\documentclass[fleqn,usenatbib]{mnras}

\usepackage{newtxtext,newtxmath}

\usepackage[T1]{fontenc}

\DeclareRobustCommand{\VAN}[3]{#2}
\let\VANthebibliography\thebibliography
\def\thebibliography{\DeclareRobustCommand{\VAN}[3]{##3}\VANthebibliography}

\usepackage{graphicx}	
\usepackage{amsmath}	

\newcommand{\cii}{{[}C\textsc{\,ii}{]}}
\newcommand{\ci}{{[}C\textsc{\,i}{]}}
\newcommand{\avg}[1]{\left\langle#1\right\rangle}

\newcommand{\dd}{\mathop{}\!{\mathrm{d}}}

\newif\iftrack
\iftrack
\newcommand{\added}[1]{{\bf #1}}
\newcommand{\deleted}[1]{}
\newcommand{\replaced}[2]{{\bf #2}}
\else
\newcommand{\added}[1]{{#1}}
\newcommand{\deleted}[1]{}
\newcommand{\replaced}[2]{{#2}}
\fi
\newif\iftracktwo
\iftracktwo
\newcommand{\addedtwo}[1]{{\bf #1}}
\newcommand{\deletedtwo}[1]{}
\newcommand{\replacedtwo}[2]{{\bf #2}}
\else
\newcommand{\addedtwo}[1]{{#1}}
\newcommand{\deletedtwo}[1]{}
\newcommand{\replacedtwo}[2]{{#2}}
\fi

\title[LIM--cosmic shear cross-correlation forecasts]{Cross-correlations between mm-wave line-intensity mapping and weak lensing surveys: preliminary consideration of long-term prospects}

\author[D. T. Chung]{
Dongwoo T.~Chung$^{1,2}$\thanks{E-mail: dongwooc@cita.utoronto.ca}
\\
$^{1}$Canadian Institute for Theoretical Astrophysics, University of Toronto, 60 St. George Street, Toronto, ON M5S 3H8, Canada\\
$^{2}$Dunlap Institute for Astronomy and Astrophysics, University of Toronto, 50 St. George Street, Toronto, ON M5S 3H4, Canada
}

\date{Accepted XXX. Received YYY; in original form ZZZ}

\pubyear{2022}

\begin{document}
\label{firstpage}
\pagerange{\pageref{firstpage}--\pageref{lastpage}}
\maketitle

\begin{abstract}
The field of millimetre-wave line-intensity mapping (LIM) is seeing increased experimental activity with pathfinder surveys already deployed or deploying in the next few years, making spectroscopic measurements of unresolved atomic and molecular line emission tracing the large-scale structure of the Universe. The next decade will also see the Rubin Observatory Legacy Survey of Space and Time (LSST) undertake a photometric galaxy survey programme of unprecedented scope, including measurements of cosmic shear exploiting weak gravitational lensing (WL) of background galaxies to map projected large-scale structure. We consider prospects for detecting angular cross power spectra between non-tomographic cosmic shear and mm-wave LIM surveys that measure emission from CO lines at $z=0.5$--1. We forecast that once the LSST Year 10 WL dataset is available, a future LIM experiment, conceivably deployed in the next 10--15 years, would enable such a cross-correlation detection with an overall signal-to-noise ratio of $50$, although the current pathfinder generation of CO/\cii{} surveys are more likely to achieve a marginal $2\sigma$ detection against an earlier-stage LSST WL dataset. The signal has modest astrophysical constraining power yielding competitive constraints on cosmic molecular gas density at $z\lesssim1$, and degeneracies between astrophysical parameters and the intrinsic alignment amplitude mean that external information on either one could allow the cross-correlation analysis to significantly improve its constraints on the other.
\end{abstract}




\section{Introduction}

Line-intensity mapping (LIM) is a nascent observational paradigm where the primary observable is not distributions of discrete resolved sources, but rather the three-dimensional fluctuations across large comoving cosmic volumes in aggregate emission in a particular spectral line. The fluctuations trace populations of unresolved galaxies and thus the underlying large-scale structure across the surveyed volume. While work to develop this technique initially centred around using emission in the 21 cm neutral hydrogen line to probe reionisation topology~\citep{Madau97,Chang08}, recent community reports~\citep{Kovetz2017,Kovetz2019,Karkare22} show significant activity in centimetre- to millimetre-wave LIM targeting carbon monoxide (CO) and ionised carbon (\cii{}) line emission.

Not only is there a significant body of literature over the past decade forecasting CO and \cii{} signals~\citep{Lidz11,Gong12,Pullen13,Mashian15,Silva15,Yue15,Li16,LidzTaylor16,Serra16,Breysse17,Padmanabhan2018a,Sun16,Bernal19b,Bernal19a,BreysseAlexandroff19,Dumitru19,Ihle19,MoradinezhadKeating19,Padmanabhan2019,Sun19,Yang21,Yang21b,Karoumpis22,MoradinezhadDizgah22a,MoradinezhadDizgah22b}, but experimental and observational activity has also begun in earnest. Early projects like the CO Power Spectrum Survey~\citep[COPSS;][]{COPSS} and the mm-wave Intensity Mapping Experiment~\citep[mmIME;][]{mmIME-ACA} making use of pre-existing interferometers have provided promising initial small-scale measurements, and the first results from a dedicated single-dish CO LIM instrument recently came from the CO Mapping Array Project~\citep[COMAP;][]{Cleary21}. Ground-based \cii{} projects like the `CarbON \cii{} line in post-rEionisation and ReionisaTiOn' experiment~\citep[CONCERTO;][]{CONCERTO}, the Tomographic Ionised-carbon Mapping Experiment~\citep[TIME;][]{Crites14}, the CCAT-prime collaboration's Epoch-of-Reionisation Spectrometer (EoR-Spec) Deep Spectroscopic Survey (DSS) on the Fred Young Submillimetre Telescope~\citep[FYST;][]{FYST}, and the South Pole Telescope Summertime Line Intensity Mapper~\citep[SPT-SLIM;][]{SPT-SLIM} have either already seen first light or will see first light in the next several years, while balloon missions like the Terahertz Intensity Mapper~\citep[TIM;][]{TIM} and the EXperiment for Cryogenic Large-Aperture Intensity Mapping~\replaced{\citep[EXCLAIM;][]{EXCLAIM}}{\citep[EXCLAIM;][]{EXCLAIM,EXCLAIMnew}} are also in development for higher-frequency and thus lower-redshift \cii{} measurements. We can expect LIM experimental programmes to continue to grow in scope and size over the coming decade well beyond this initial generation of pathfinder experiments.

Meanwhile, it is common knowledge that the scope and size of galaxy surveys has been growing steadily over the past century. The culmination of photometric imaging surveys for the foreseeable future will be the Vera C.~Rubin Observatory Legacy Survey of Space and Time~\citep[LSST;][]{LSST}, which will tackle a wide range of transient, solar, Galactic, and extragalactic science in a programme spanning ten years and 18000 deg$^2$. In particular, as part of its cosmological science operations, LSST will make use of weak gravitational lensing (WL) techniques, where intervening large-scale structure distorts the shapes of background galaxies in a correlated way, dubbed cosmic shear. Measurements of cosmic shear thus probe the projected distribution of matter between us and the distorted background sources, and already demonstrate cosmological constraining power in recent results from the Subaru Hyper Suprime-Cam (HSC) survey~\citep{Hikage19}, the Kilo-Degree Survey~\citep[KiDS;][]{Heymans21}, and the Dark Energy Survey~\citep[DES;][]{Doux22}.

Since LIM and cosmic shear measurements both probe the same cosmic web, cross-correlations between the two would not only validate one measurement of large-scale structure against another, but also yield synergistic science output.\added{ Similar cross-correlations have been considered and measured in entirely non-spectroscopic contexts, such as between cosmic microwave background lensing convergence and cosmic infrared background imagery~\citep[e.g.,][]{Holder13}.} Previous work has in fact considered the potential of cross-correlating LIM data against WL data~\citep[e.g.:][]{Foreman18,SchaanWhite21,Shirasaki21}. However, \cite{Foreman18} examined cross-correlation of cosmic shear against lensing of the line-intensity field, rather than against the line-intensity field itself. \cite{Shirasaki21} examine cross-correlation of cosmic shear against the LIM observation rather than its lensing, but in the context of constraining decaying dark matter (and outside the context of mm-wave LIM, where~\cite{Bernal21} indicate constraints will not necessarily be competitive). \cite{SchaanWhite21} do discuss LIM--WL cross-correlation briefly but avoid explicit projections. Above all, the current literature does not set expectations for detectability of and constraints from direct correlation of WL data with the actual astrophysical lines targeted by LIM experiments. But CO lines from $z=0.5$--1 are an excellent target for such cross-correlation, being more visible in sum than the high-redshift \cii{} line in mm-wave LIM data and originating from redshifts where the WL kernel peaks. To motivate further future work, we show here that a cross-correlation between these CO lines and cosmic shear is possible with future data, and potentially of scientific value.

This work aims to answer, on a strictly preliminary basis, the following questions:
\begin{itemize}
    \item What is the expected detection significance of cosmic shear--LIM cross-correlation?
    \item What kinds of quantities could this cross-correlation constrain, astrophysical or otherwise?
\end{itemize}

We organise the paper as follows. We first provide an overview of the experimental context in~\autoref{sec:expcontext} before establishing models for LIM and WL observables and power spectra in~\autoref{sec:models}. Then we forecast signal detectability and some possible model parameter constraints in~\autoref{sec:forecasts} before finally concluding in~\autoref{sec:conclusions}.

Throughout this work, we assume base-10 logarithms unless otherwise stated, and a $\Lambda$CDM cosmology with parameters $\Omega_\mathrm{m} = 0.307$, $\Omega_\mathrm{b} =0.0486$, $H_0=100h$\,km\,s$^{-1}$\,Mpc$^{-1}$ with $h=0.677$, $\sigma_8 =0.8159$, and $n_s =0.9667$, consistent with 2015 cosmological parameter constraints from the \emph{Planck} satellite mission~\citep{Planck15}. Distances carry an implicit $h^{-1}$ dependence throughout, which propagates through masses (all based on virial halo masses, proportional to $h^{-1}$) and volume densities ($\propto h^3$).

\section{Experimental parameters}
\label{sec:expcontext}

Here we briefly describe the parameters that we need to define for LIM and WL experiments in order to determine their respective uncertainties, and decide on two fiducial scenarios for cross-correlation. 

\subsection{Cosmic shear: LSST}
We consider two scenarios for WL data, both based on the LSST Dark Energy Science Collaboration (DESC) Science Requirements Document~\citep[SRD, version 1.0.2;][]{LSST-SRD}. The LSST DESC SRD refers throughout to forecasts for `Year 1' (Y1) and `Year 10' (Y10) datasets, corresponding to expected survey depths and areas achieved with one-tenth of LSST observations and with all observations from the ten-year campaign. At the time of writing, Rubin Observatory is scheduled to begin LSST operations in the first half of 2024~\citep{RubinEarlySci}, meaning Y10 will encompass data taken through the end of 2033. The DESC SRD takes care to note, however, that a Y1-like dataset will not necessarily be deliverable one year after the start of survey operations, and indeed the timeline for achieving the simulated survey depths will depend strongly on the observing strategy.

For both Y1 and Y10 datasets, the expected sky coverage is in excess of $10^4$ deg$^2$, which in turn is well in excess of the sky coverage we will consider for our mm-wave LIM concepts. Then for our cross-correlation scenario, the important difference between the Y1 and Y10 WL datasets is the redshift distribution and number density of the source galaxy samples that are distorted along the cosmic shear field, both outlined in Appendix D of the DESC SRD. The expected effective source number densities $n_g$ for Y1 and Y10 data are respectively 10 arcmin$^{-2}$ and 27 arcmin$^{-2}$. The cosmic shear shot noise also depends on the standard deviation of measured ellipticity per component, held to be $\sigma_e=0.26$ in all cases.

The DESC SRD parameterises the source galaxy redshift probability distribution $n(z)$ as follows:
\begin{equation}
    n(z)\propto z^2\exp{\left(-\frac{z}{z_0}\right)^{\alpha_\text{src}}}.
\end{equation}
(We have added a descriptive subscript to $\alpha_\text{src}$ as other parameters in this work will be named $\alpha$ with various subscripts or no subscript.) We normalise $n(z)$ such that $\int n(z)\,dz = 1$, as befits a probability distribution. Because of the increase in survey depth between Y1 and Y10 datasets, the DESC SRD expects the parameters for $n(z)$ to shift from $(z_0,\alpha_\text{src})=(0.13,0.78)$ to $(z_0,\alpha_\text{src})=(0.11,0.68)$, resulting in an upward shift in the median source redshift.

\subsection{Millimetre-wave LIM: FYST DSS and future concepts}
Since LIM is a survey of aggregate emission rather than individual galaxies, the instrument noise fully determines the uncertainty of the LIM measurement (excluding sample variance). At millimetre wavelengths, detectors are within striking distance of background-limited operation, meaning that for forecasting purposes we may assume that the dominant source of system emissivity and thus noise is loading from the atmosphere.

We will focus on the 200--300 GHz atmospheric window, staying well clear of water vapour lines centred at 183 GHz and 325 GHz. It is worth noting that higher frequencies are observable, especially at extremely dry sites like at the South Pole or at high altitudes in the Atacama desert, and indeed EoR-Spec DSS on FYST is forecast to make \cii{} detections at observing frequencies of 350 GHz and above~\citep{FYST}. Nonetheless, for the purposes of this work we will focus on a narrower observing band for the sake of simplicity.

Based on Appendix B of~\cite{cii_um}, the noise-equivalent intensity per detector scales with channel bandwidth $\delta\nu$ and system emissivity $\epsilon_\text{sys}$ as follows:
\begin{equation}
    \sigma_\text{spec}=10^6\text{ Jy\,sr}^{-1}\,\text{s}^{1/2}\left(\frac{\delta\nu}{2.5\,\text{GHz}}\right)^{-1/2}\left(\frac{\epsilon_\text{sys}}{0.05}\right),
\end{equation} 
under the implicit assumption that the detection bandwidth is equal to the frequency channel width, which is true for EoR-Spec but not for Fourier-transform spectrometers like the CONCERTO instrument. (The noise scales proportionally with the square root of the detection bandwidth but is inversely proportional to $\delta\nu$.) Note that when assuming a 5\% emissivity (effectively 95\% atmospheric transmission given our simplifications),~\cite{cii_um} independently calculated a noise-equivalent flux density (NEFD) per beam of 69 mJy s$^{1/2}$ for EoR-Spec at 250 GHz, in reasonable agreement with the first-quartile (median) NEFD per beam of 68 (81) mJy s$^{1/2}$ quoted for EoR-Spec at 220 GHz~\citep{FYST}.

In our calculations, we will introduce frequency-dependence in $\sigma_\text{spec}$ through both $\delta\nu$ and $\epsilon_\text{sys}$. Following specifications for EoR-Spec we assume a resolving power of $\nu/\delta\nu=100$. As for emissivity, we use expected atmospheric transmission based on the ATM model~\citep{Pardo01} to estimate the frequency-dependent $\epsilon_\text{sys}$ as varying between 4\% and 8\% across the 200--300 GHz band from Chajnantor Plateau, assuming 0.6 mm precipitable water vapour (PWV) at zenith and source elevation of 45 degrees.\footnote{We use a calculator provided by the Atacama Pathfinder EXperiment (APEX) website at \url{https://www.apex-telescope.org/ns/weather/}.}

For reference, the first-quartile (median) PWV is 0.36 (0.67) mm across the year at the Cerro Chajnantor summit just above the FYST site~\citep{Cortes20}. Note however that atmospheric transmission for a given PWV value depends strongly on site characteristics. At the South Pole Telescope (SPT) site, a study for the Event Horizon Telescope collaboration by~\cite{Raymond21} found median PWV to fluctuate between 0.24 mm in July and 0.8 mm in January, but the median (75th percentile) zenith opacity at 230 GHz for each month never rose above 6\% (7\%) across the whole year. The upshot is that the level of atmospheric loading that we assume for LIM sensitivities is readily achievable from locations like the FYST and SPT sites, even easily so depending on the time of year.

With $\sigma_\text{spec}$ defined, we still need to define the integration time per solid angle per spectral channel. This should be given simply by the survey time $t_\text{surv}$ multiplied by the spectrometer count---a product that we will refer to as the number of spectrometer-hours, borrowing a choice of vocabulary from~\cite{MoradinezhadDizgah22a}---and divided by the solid angle of the survey area. The EoR-Spec DSS concept prescribes coverage of two 4 deg$^2$ fields; we make a simplification and make forecasts for a continuous 8 deg$^2$ field (the resulting difference in sensitivity being a factor of order unity). While the~\cite{FYST} parameters quote a detector count of $6348$ in the 210--315 GHz band, this is before a possible 80\% yield factor, and even then the Fabry-Perot interferometer must scan across 42 different settings for each detector to obtain a full spectrum. Therefore, we consider EoR-Spec to effectively comprise $N_\text{spec,eff}\approx6348\cdot0.8/42\approx120$ spectrometers. Since the total survey time across the 8 deg$^2$ sky coverage is $t_\text{surv}=4000$ hours, this places the effective integration time for EoR-Spec DSS on FYST at $4.8\times10^5$ spectrometer-hours across the survey area.

What remains then is a plausible spectrometer-hour count for a future experiment that would plausibly deploy towards the end of the LSST campaign and be available for cross-correlation with a Y10-like dataset. \cite{Karkare22} outline a possible path for future stages of mm-wave LIM experiments to scale up in spectrometer-hours. If we treat current experiments like TIME with $\sim10^5$ spectrometer-hours as a sort of Stage 0 for mm-wave LIM, then under their approximate timeline, a mm-wave Stage 3 ($\sim10^8$ spectrometer-hour) experiment could conceivably begin survey operations in the early 2030s, around the same time that LSST would near the end of its ten-year survey programme. Therefore, we consider a Stage 3 concept with $1.8\times10^8$ spectrometer-hours covering 1024 square degrees of sky, in cross-correlation with a LSST Y10-like WL dataset. With 375 times the total survey integration time over 128 times the sky area, such a survey would still result in a factor-of-3 improvement in map noise and thus almost an order-of-magnitude increase in the LIM auto spectrum detection significance compared to FYST DSS forecasts. We assume conservatively that other parameters like channel bandwidth and system emissivity remain the same between EoR-Spec and this Stage 3 experiment.

Before moving on to signal models, we resummarise the LIM and WL parameters we have discussed so far in~\autoref{tab:expcontext}. There we also show the LIM survey area as a fraction $f_\text{sky}$ of all $4\pi$ sr or $\approx41253$ deg$^2$ of the celestial sphere.

\begin{table*}
    \centering
    \begin{tabular}{cccccccc}
         \hline Cross-correlation scenario&\multicolumn{3}{c}{LSST-like source distribution:} & \multicolumn{4}{c}{LIM survey parameters:}\\
         &$\alpha_\text{src}$&$z_0$&$n_\text{g}$&Survey area&$f_\text{sky}$&\added{$f_\text{sky}\times100$\%}&$N_\text{spec,eff}t_\text{surv}$\\
         && & (arcmin$^{-2}$)& (deg$^2$)&&&(hours)\\\hline
         LSST Y1-like $\times$ FYST DSS-like& 0.78 & 0.13 & 10 & 8 & $1.9\times10^{-4}$ & \added{0.019\%} & $4.8\times10^5$\\
         LSST Y10-like $\times$ mm-wave Stage 3 & 0.68 & 0.11 & 27 & 1024 & $2.4\times10^{-2}$ & \added{2.4\%} & $1.8\times10^8$\\\hline
    \end{tabular}
    \caption{Lensing and LIM survey parameters varied in different experimental scenarios.}
    \label{tab:expcontext}
\end{table*}

\section{Models: observables and statistics}
\label{sec:models}
\subsection{Line-intensity fields}
\label{sec:linemodels}

The mm-wave LIM surveys under consideration will observe \cii{}, \ci{}, and CO line emission, all of which are associated with star-formation activity. Therefore, we follow the approach of various prior works~\citep{Lidz11,Gong12,Pullen13,Li16,Silva15,Serra16,Sun16,Sun21} in formulating a halo model that first connects halo mass and redshift to an expected value for star-formation rate (SFR), then ties this to line luminosities via observational fits relating to SFR proxy luminosities.

\subsubsection{Halo mass to SFR: \textsc{UniverseMachine} DR1}
Although some prior works rely on halo models of the cosmic infrared background to calibrate the halo mass--SFR relation~\citep{Serra16,Sun16,Sun21}, our approach will tie closer to that of~\cite{Li16} and~\cite{cii_um} in using empirical halo models of galaxy evolution. These models, like that of~\cite{Behroozi13a,Behroozi13b} as used by~\cite{Li16}, are calibrated simultaneously against halo merger and mass accretion histories in cosmological simulations and against observed cosmic stellar mass and star-formation rate densities, although we will not be working with simulation lightcones. We specifically use the \textsc{UniverseMachine} DR1 best-fit model from~\cite{Behroozi19}, which with certain simplifying assumptions can be made to provide a reasonable approximate halo mass--SFR relation.

The average SFR for star-forming galaxies is given by a double power-law relation with Gaussian enhancement near the characteristic mass:
\begin{equation}
    \avg{\mathrm{SFR_{SF}}}=\epsilon\left[\frac{1}{v^\alpha+v^\beta}+\gamma\exp{\left(-\frac{\log^2{v}}{2\delta^2}\right)}\right],
\end{equation}
where $v=v_{M_\text{peak}}/V$ (with $V$ in units of km s$^{-1}$) is the maximum circular velocity of the halo at peak mass divided by the model parameter $V$, and $\gamma$, $\delta$, and $\epsilon$ are all model parameters, with the first two describing the Gaussian enhancement. We refer the reader to~\cite{Behroozi19} for parameter values, as each parameter is itself controlled by several parameters due to functional forms with nontrivial redshift dependence.

This maximum velocity $v_{M_\text{peak}}$ is approximately related to the halo mass $M_h$ by a median relation based on the Bolshoi-Planck simulation provided in Appendix E2 by~\cite{Behroozi19}:
\begin{equation}
    v_{M_\text{peak}}(M_h) = (200\text{ km s}^{-1})\left[\frac{M_h}{M_\text{200 km/s}(a)}\right]^{0.3},
\end{equation}
where in turn
\begin{equation}
    M_\text{200 km/s} = \frac{1.64\times10^{12}\,M_\odot}{(a/0.378)^{-0.142}+(a/0.378)^{-1.79}},
\end{equation}
with a double power-law dependence on the scale factor $a=1/(1+z)$.

The~\cite{Behroozi19} best-fit model gives the quenched galaxy fraction as
\begin{align}
    f_Q &= Q_\text{min} + \frac{1-Q_\text{min}}{2}\left[1+\operatorname{erf}{\left(\frac{\log{\left(v_{M_\text{peak}}/(V_Q\cdot\text{km s}^{-1})\right)}}{\sigma_{VQ}\sqrt{2}}\right)}\right],
\end{align}
controlled by the model parameters $Q_\text{min}$, $V_Q$, and $\sigma_{VQ}$ (for which we refer the reader again to~\citealt{Behroozi19}). Then, per Equation E4 of~\cite{Behroozi19}, the average SFR for a given halo mass at redshift $z$ is
\begin{equation}
    \avg{\mathrm{SFR}}(M_h,z) \approx \frac{1-f_Q}{C_\sigma}\avg{\mathrm{SFR}_\text{SF}}(v_{M_\text{peak}}(M_h),z),
\end{equation}
with $C_\sigma$ encapsulating a slight multiplicative discrepancy due to scatter in the relations between $M_h$, $v_{M_\text{peak}}$, and SFR. For our purposes, we take $C_\sigma=2/3$, which appears to provide a reasonable match to model average SFR and cosmic SFR density values in~\cite{Behroozi19}.

\subsubsection{SFR to line luminosities}

To convert SFR to atomic and molecular lines, we use prescriptions similar to those of~\cite{Sun21}, which ties line luminosities to either UV or IR luminosity, both ultimately connected to SFR.

A 200--300 GHz survey will measure \cii{} line emission (rest frequency $\nu_\text{rest,\cii{}}=1901.0302$ GHz) from ionised carbon between redshifts 5.3 and 8.5. \cite{Sun21} assumes that \cii{} luminosity scales linearly with UV continuum luminosity $L_\text{UV}$. We substitute an assumed proportionality between $L_\text{UV}$ and SFR:
\begin{align}
    \log\left(\frac{L_\text{\cii{}}}{L_\odot}\right) &= \log{\left(\frac{L_\text{UV}}{\text{erg s}^{-1}\text{ Hz}^{-1}}\right)}-20.6\nonumber\\&= \log{\left(\frac{\avg{\mathrm{SFR}}(M_h,z)}{1.15\times10^{-28}\,M_\odot\text{ yr}^{-1}}\right)}-20.6\nonumber\\&= \log{\left(\frac{\avg{\mathrm{SFR}}(M_h,z)}{M_\odot\text{ yr}^{-1}}\right)}+7.34.
\end{align}
The value of the UV--SFR conversion factor depends specifically on assuming a Salpeter initial mass function (IMF) and originates from~\cite{SunFurlanetto16}.

The procedure for CO lines, which arise from a ladder of rotational energy transitions, is similar. We index these lines based on the rotational quantum number $J$ corresponding to the upper level involved in the $J\to J-1$ transition, with the CO$(J\to J-1)$ line having a rest emission frequency of $\nu_{\text{rest,CO}(J\to J-1)}=\nu_\text{rest,CO(1--0)}J=(115.27\cdot J)$ GHz.\footnote{This is technically not quite true. As~\cite{ERA} note, the rotational moment of inertia of the CO molecule increases with higher rotational energy due to centrifugal stretching of the diatomic bond, leading to $\nu_{\text{rest,CO}(J\to J-1)}$ not scaling perfectly linearly with $J$. However, for our purposes $\nu_{\text{rest,CO}(J\to J-1)}$ is close enough to $(115.27\cdot J)$ GHz.}

As with the UV luminosity, we assume that the IR luminosity $L_\text{IR}$ scales linearly with SFR, this time taking our conversion factor from~\cite{Sun19} (which again assumes a Salpeter IMF):
\begin{equation}
    \frac{\avg{\mathrm{SFR}}(M_h,z)}{L_\text{IR}}=\frac{1.73\times10^{-10}\,M_\odot\text{ yr}^{-1}}{L_\odot}.
\end{equation}

Observers often express CO line luminosities as velocity- and area-integrated brightness temperature, and the conversion between this quantity and intrinsic luminosity is commonly used by works like~\cite{Li16} and~\cite{Sun21}:
\begin{equation}
    \frac{L_\text{line}}{L_\odot} = 4.9\times10^{-5}\left(\frac{\nu_\text{rest,line}}{115.27\text{ GHz}}\right)^3\frac{L'_\text{line}}{\text{K km s}^{-1}\text{ pc}^2}.
\end{equation}
We have expressed this conversion so that for the CO lines, the cubic factor depending on rest frequency becomes simply $J^3$.

All CO line luminosities are assumed to scale linearly with the CO(1--0) line luminosity, which~\cite{Sun21} relate to IR luminosity via a common power-law functional form:
\begin{align}
    \log\left(\frac{L'_\text{CO(1--0)}}{\text{K km s}^{-1}\text{ pc}^2}\right) &= \alpha_\text{IR--CO}^{-1}\left[\log{\left(\frac{L_\text{IR}}{L_\odot}\right)}-\beta_\text{IR--CO}\right]\nonumber\\&= \frac{\log{\left(\frac{\avg{\mathrm{SFR}}(M_h,z)}{M_\odot\text{ yr}^{-1}}\right)}+9.76-\beta_\text{IR--CO}}{\alpha_\text{IR--CO}}.\label{eq:LCO10}
\end{align}
To obtain the scaling factors $r_J$ between CO(1--0) and CO($J\to J-1$) line luminosities,~\cite{Sun21} use a \emph{Herschel}/SPIRE analysis by~\cite{Kamenetzky16} of CO excitation and the CO spectral line energy distribution (SLED). \cite{Sun21} also cite~\cite{Kamenetzky16} for fiducial values of $\alpha_\text{IR--CO}=1.27$ and $\beta_\text{IR--CO}=-1.0$ for the CO--IR luminosity relation.

Ultimately, the relation between the SFR and CO line luminosities is as follows:
\begin{align}
    \log\left(\frac{L_{\text{CO}(J\to J-1)}}{L_\odot}\right) &= \alpha_\text{IR--CO}^{-1}\left[\log{\left(\frac{\avg{\mathrm{SFR}}(M_h,z)}{M_\odot\text{ yr}^{-1}}\right)}+9.76\right.\nonumber\\&\qquad\qquad\qquad\left.{}-\beta_\text{IR--CO}\right]-4.31+\log{(r_JJ^3)},
\end{align}
where $r_3=0.73$, $r_4=0.57$, $r_5=0.32$, $r_6=0.19$, and $r_7=0.1$.

\cite{Sun21} also consider \ci{} line emission ($\nu_\text{rest,\ci{}}=492.16$ GHz) from neutral carbon in a similar manner, noting that the critical density and thus excitation of \ci{} is similar to that of CO (cf.~the \citealt{CW13} review of high-$z$ molecular gas). They also cite modelling works~\citep{Bisbas15,Glover15} that potentially explain \ci{}--CO correlations, and observational literature~\citep{Israel15,Jiao17,Valentino18,Nesvadba19} using \ci{} to trace molecular gas at low and high redshifts. In particular, \cite{Sun21} cite~\cite{Jiao17} as an example of a study that finds a linear correlation between \ci{} and CO(1--0) luminosities, using this finding to effectively model the \ci{} line in the same way as the CO lines:
\begin{align}
    \log\left(\frac{L_{\text{\ci{}}}}{L_\odot}\right) &= \alpha_\text{IR--CO}^{-1}\left[\log{\left(\frac{\avg{\mathrm{SFR}}(M_h,z)}{M_\odot\text{ yr}^{-1}}\right)}+9.76-\beta_\text{IR--CO}\right]\nonumber\\&\qquad-4.31+\log{(r_\text{\ci{}}J_\text{\ci{}}^3)},
\end{align}
where $r_\text{\ci{}}=0.18$ and $J_\text{\ci{}}\equiv\nu_\text{rest,\ci{}}/\nu_\text{rest,{CO(1--0)}}=4.27$. \cite{Sun21} calibrate their determination of $r_\text{\ci{}}$ against \ci{}--IR luminosity relations found in the literature (specifically citing~\citealt{Valentino18} and~\citealt{Nesvadba19}).

\subsubsection{Integrated quantities}
We use certain quantities integrated across the halo mass function when calculating LIM auto- and cross-correlation signals. These are the average line intensity $\avg{I}_\text{line}$, the average line intensity-bias product $\avg{Ib}_\text{line}$, and the three-dimensional shot noise power spectrum $P_\text{shot,line}$. $\avg{I}_\text{line}$ and $P_\text{shot,line}$ respectively reflect the first and second moments of the line-luminosity function, with clustering-scale line-intensity fluctuations tracing matter fluctuations with linear bias $\avg{Ib}_\text{line}/\avg{I}_\text{line}$.

For the sake of readability in calculating these quantities, we adopt notation similar to~\cite{BreysseAlexandroff19} and define a conversion factor $C_\text{LI}$ between luminosity density and line intensity for a line with rest frequency $\nu_\text{rest,line}$ emitted at redshift $z$:
\begin{align}
    C_\text{LI}(\nu_\text{rest,line},z)=\frac{c}{4\pi\nu_\text{rest,line}H(z)},
\end{align}
with $H(z)$ being the Hubble parameter at redshift $z$.

We use \texttt{hmf}~\citep{hmf} to obtain values for the halo mass function based on the model of~\cite{Tinker08}, requiring a minimum halo mass of $10^{10}h^{-1}\,M_\odot$ for nonzero line emission. We also calculate the linear halo bias $b(M_h)$ following the model given by~\cite{Tinker10}. Then the three integrated quantities discussed above are obtained from the halo mass function $dn/dM_h$ as follows:
\begin{align}
    \avg{I}_\text{line} &= C_\text{LI}(\nu_\text{rest,line},z)\int \dd{M_h}\,\frac{\dd{n}}{\dd{M_h}}\,L_\text{line}(M_h,z);\\
    \avg{Ib}_\text{line} &= C_\text{LI}(\nu_\text{rest,line},z)\int \dd{M_h}\,\frac{\dd{n}}{\dd{M_h}}\,L_\text{line}(M_h,z)b(M_h,z);\\
    P_\text{shot,line} &= C_\text{LI}^2(\nu_\text{rest,line},z)\int \dd{M_h}\,\frac{\dd{n}}{\dd{M_h}}\,L^2_\text{line}(M_h,z)\nonumber\\&\qquad\times\exp{\left[(\sigma_\text{line}\ln{10})^2\right]},
\end{align}
where in defining $P_\text{shot,line}$ we have allowed for a possible enhancement of shot noise due to log-normal halo-to-halo scatter around the average halo mass--line luminosity relation, with standard deviation $\sigma_\text{line}$ in units of dex. Throughout the remainder of this work we will assume $\sigma_\text{line}=0.4$ dex for all lines and take this scatter to be perfectly correlated between all lines.

While the most relevant quantities for power spectra are $\avg{Ib}_\text{line}$ and $P_\text{shot,line}$, $\avg{I}_\text{line}$ is also useful to calculate because it is directly proportional to the line luminosity density and thus to quantities like the molecular gas density $\rho_\text{H2}$ that scale linearly with line luminosity densities. For reference, we show $\avg{I}_\text{line}$ at a mid-band frequency of 251 GHz, which are within a factor of order unity of the predictions of~\cite{Sun21} at 250 GHz. The difference that does exist between the two sets of predictions is certainly mostly due to the differences in the $\avg{\mathrm{SFR}}(M_h,z)$ relation used\added{, and may be considered an approximate lower-bound characterisation of the level of uncertainty around the astrophysical signals being modelled}.

\begin{table}
    \centering
    \begin{tabular}{cccc}
         \hline Line & Redshift & $\avg{I}_\text{line}$ predicted & Ratio to\\
         & at 251 GHz & at 251 GHz & \cite{Sun21}\\
         && (Jy sr$^{-1}$) & prediction\\\hline
         CO(3--2) & 0.38 & 188 & 0.80\\
         CO(4--3) & 0.84 & 396 & 0.77\\
         \ci{} & 0.96 & 151 & 0.77\\
         CO(5--4) & 1.30 & 400 & 0.73\\
         CO(6--5) & 1.76 & 328 & 0.68\\
         CO(7--6) & 2.21 & 203 & 0.64\\
         \cii{} & 6.57 & 140 & 0.36\\\hline
    \end{tabular}
    \caption{Average line intensities mid-band as predicted by the model of~\autoref{sec:linemodels}, and their ratios to predictions of~\protect\cite{Sun21}.}
    \label{tab:Ilines}
\end{table}

\subsection{WL fields}

In modelling the WL cosmic shear measurement, we largely follow work laid out very briefly by~\cite{SchaanWhite21} and in more depth by~\cite{Kilbinger15} and~\cite{Shirasaki21}. The cosmic shear measurement will have three components: the actual cosmic shear field from intervening large-scale structure between the observer and the source galaxies, the intrinsic alignment of galaxy shapes along large-scale structure underlying the source galaxies, and a shot noise component due to intrinsic ellipticity variance (which we discussed briefly in~\autoref{sec:expcontext}). There is not much to discuss about the last component, which is fully determined by the parameters $n_g$ and $\sigma_e=0.26$ discussed in~\autoref{sec:expcontext}. However, we will discuss our models for cosmic shear and intrinsic alignments in this section.
\subsubsection{Cosmic shear}
We work in the Born approximation, which \cite{Kilbinger15} expresses as working with a zeroth-order approximation to the separation vector (ignoring lensing deflections), and a linearised mapping from lensed to unlensed coordinates. All of this is perfectly valid in the WL regime where sources are not lensed into multiple images and distortions are percent-level.

The linearised mapping is fully described by the lensing convergence $\kappa$ and a two-component shear often expressed as a complex number $\gamma$, respectively describing isotropic magnification and anisotropic distortion. The cosmic shear measurement is a measurement of shape distortion, not magnification, with deviation of observed ellipticity from intrinsic ellipticity being an unbiased estimator of reduced shear $\gamma/(1-\kappa)\approx\gamma$. However, both convergence and shear are related to second derivatives of the 2D lensing potential, and the convergence in particular encapsulates the projected surface density of the distorting large-scale structure. This is why we end up describing the cosmic shear measurement in terms of $\kappa$.

We refer the reader to \cite{Kilbinger15} for a more detailed review, but in short the lensing kernel scales matter density contrast at comoving distance $\chi$ based on the distribution $n(z)$ of source galaxies:
\begin{equation}
    W_\kappa(\chi)=\frac{3H_0^2\Omega_\mathrm{m}}{2c^2}(1+z)\int_\chi^{\chi_\text{lim}} \dd{\chi'}\,\frac{\dd{z'}}{\dd{\chi'}}n(z')\frac{\chi(\chi'-\chi)}{\chi'},
\end{equation}
such that the convergence field is $\kappa(\boldsymbol{\theta})=\int \dd{\chi}\,W_\kappa(\chi)\,\delta_\mathrm{m}(\chi\boldsymbol{\theta},\chi)$ given the matter density contrast field $\delta_\mathrm{m}$. Here $z$ and $z'$ are redshifts respectively corresponding to the $\chi$ and $\chi'$, and $\chi_\text{lim}$ is the limiting comoving distance of the galaxy sample. Being conservative, we take $\chi_\text{lim}=\chi(z=20)$ to avoid underestimating potential outlier galaxies at high redshift in the LSST source sample (which the DESC SRD notes has no upper redshift cutoff). Recall also that in this work, we have normalised $n(z)$ such that $\int \dd{z}\,n(z)=1$.

\subsubsection{Intrinsic alignments}
\label{sec:IAmodel}
Cosmic shear measurements depend on measuring the ellipticities of a population of source galaxies, but these sources are also part of the cosmic web and interact gravitationally with it. The tidal interactions of source galaxies with large-scale structure (and thus with each other to some extent) result in an intrinsic alignment (IA) of galaxy shapes, resulting not only in correlation of shapes between nearby galaxies but also correlation of ellipticity with shear. All of this adds excess correlations to the observed shear signal. The correlations are orthogonal to those expected from WL, however, due to the expected radial alignment of galaxy shapes towards matter overdensities versus the tangential distortion that results in the WL cosmic shear signal.

We use a model for IA dubbed the linear alignment model. This model was first proposed by~\cite{Catelan01}, explored further by~\cite{Hirata04}, and subsequently expanded on by many others including~\cite{Hirata07} and~\cite{BridleKing07}. The central assumption is that apparent ellipticities scale linearly with second derivatives of the gravitational potential (i.e., the tidal field). When altering this model to use the non-linear matter power spectrum (instead of the linear matter power spectrum) to determine IA-related terms in the observed cosmic shear power spectrum, \cite{BridleKing07} dubbed their phenomenological variation the `non-linear linear alignment' (NLA) model. Recent works (e.g.,~\citealt{Hikage19,Doux22}) however often simply refer to this variation as the non-linear alignment model, even though the alignments still scale linearly with the tidal field.

We use a form of this model used by~\cite{Hikage19} for the Subaru HSC Y1 cosmic shear analysis and restated by~\cite{Shirasaki21}. By analogy to cosmic shear, we consider that the IA effect contributes an apparent convergence field $\kappa_\text{IA}(\boldsymbol{\theta})=\int \dd{\chi}\,W_\text{IA}(\chi)\,\delta_\mathrm{m}(\chi\boldsymbol{\theta},\chi)$, where
\begin{equation}
    W_\text{IA}(\chi)\,\dd\chi = -A_\text{IA}C_1\frac{\Omega_\mathrm{m}\rho_{\text{crit},0}}{D(z)}\left(\frac{1+z}{1+z_{0,\text{IA}}}\right)^{\eta_\text{IA}}n(z')\,\dd{z}.
\end{equation}
Here $C_1=5\times10^{-14}h^{-2}\,M_\odot^{-1}$ Mpc$^3$ (a convention based on SuperCOSMOS Sky Survey results presented by~\citealt{Brown02}), and $D(z)$ is the normalised linear growth factor such that $D(z=0)=1$. This leaves three model parameters, for which we will assume the same fiducial values as~\cite{Shirasaki21} of $z_{0,\text{IA}}=0.62$, $\eta_\text{IA}=3$, and $A_\text{IA}=1$.

The first two choices are based on fixed or plausible values discussed by~\cite{Hikage19}, but the choice of $A_\text{IA}$ is extremely speculative as various studies have found different values. The HSC Y1 analysis of~\cite{Hikage19} found a best estimate of $A_\text{IA}=0.38\pm0.70$ (although higher with alternate choices of code for photometric redshift estimation), with no firm constraint on $\eta_\text{IA}$. Working with a redshift-independent version of the NLA model, the KiDS-1000 results found a best estimate of $A_\text{IA}=1.07^{+0.27}_{-0.31}$ combining cosmic shear in a joint $3\times2$pt analysis with galaxy--galaxy lensing and galaxy clustering; results from cosmic shear band power alone presented by~\cite{Asgari21} place the best-fit value slightly but insignificantly lower at $0.973^{+0.292}_{-0.383}$. DES Y3 shear $C_\ell$ results shown by~\cite{Doux22} found $A_\text{IA}=0.40\pm0.51$ however, more consistent with the HSC Y1 result. We choose to set $A_\text{IA}=1$ as~\cite{Shirasaki21} did, mirroring the high end of the values found in observational literature and thus erring on the conservative side.

We apply this model uniformly to all galaxies, assuming that the redshift-dependent term sufficiently encapsulates the variability of intrinsic alignments across different types of galaxies. \cite{Fortuna21} suggests that this model (which they dub the NLA-$z$ model to indicate redshift-dependence) is sufficiently flexible to result in unbiased cosmological constraints without needing to model separate amplitudes for red and blue galaxies (although ultimately this will depend on the galaxy selection). While we will not be making use of this flexibility or forecasting cosmological constraints, this should indicate that the NLA (or rather NLA-$z$) model is suitable for use in this context even as more sophisticated methods have come into use, e.g., the tidal alignment and tidal torquing (TATT) model of~\cite{Blazek19} used by~\cite{Doux22}.

\subsection{Angular power spectra}
\label{sec:celloutline}
\begin{figure}
    \centering
    \includegraphics[clip=True,trim=2mm 5mm 5mm 2mm,width=0.96\linewidth]{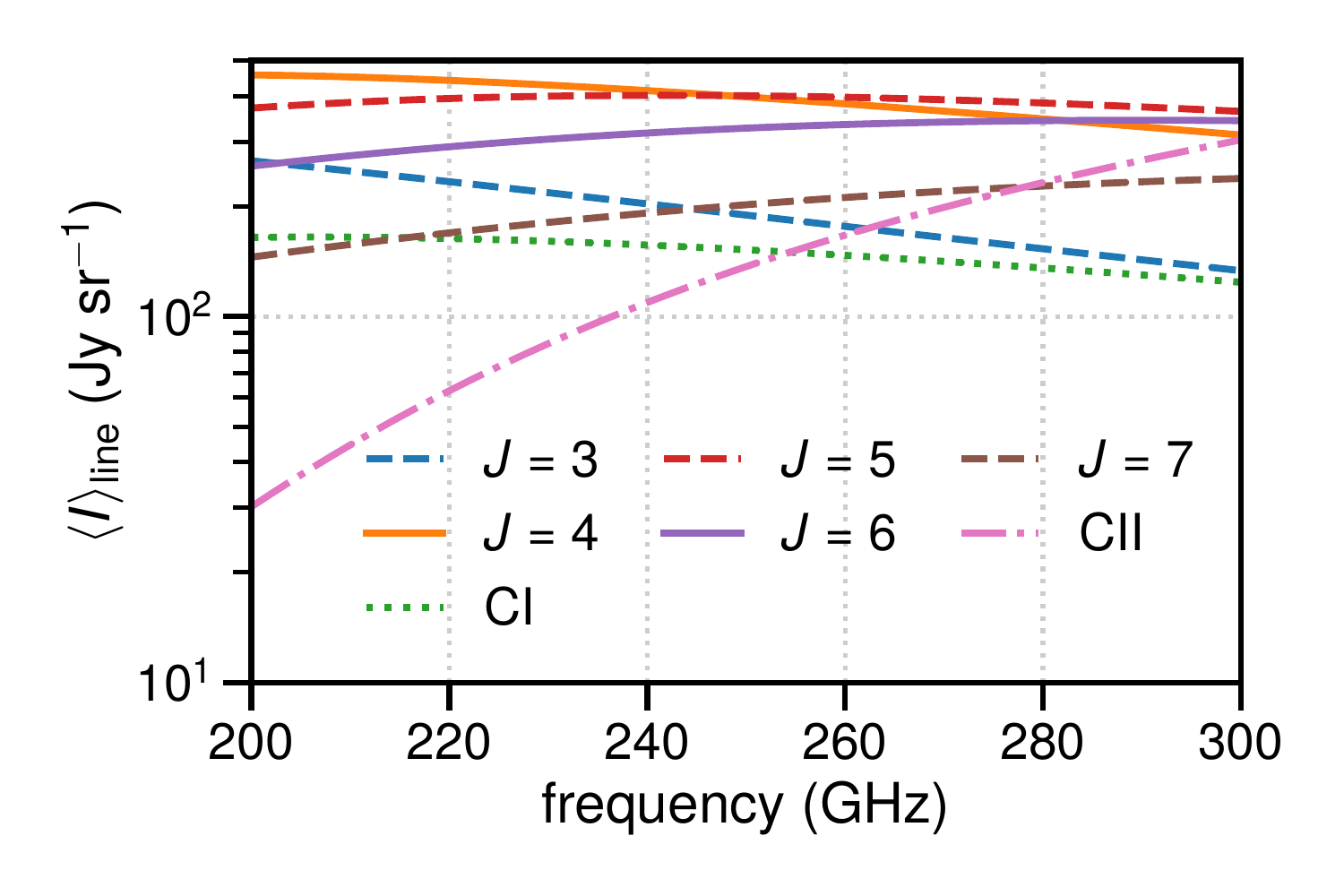}
    
    \includegraphics[clip=True,width=0.92\linewidth]{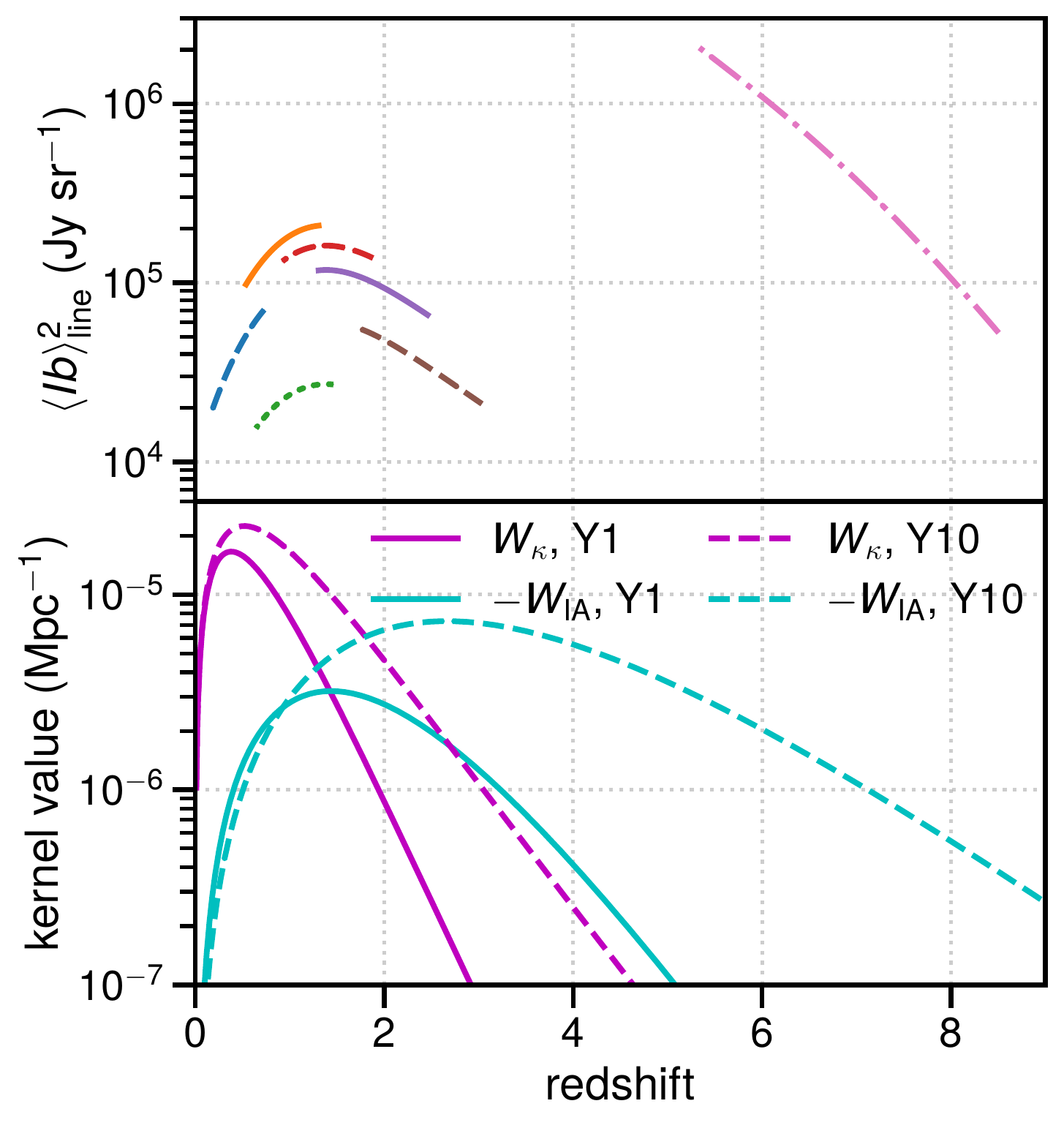}
    \caption{Illustration of observables, in the form of mean line intensity $\avg{I}_\text{line}$ for all astrophysical lines considered as a function of observing frequency (upper panel), and \replaced{lensing and IA kernels for LSST Y1- and Y10-like galaxy distributions as a function of redshift (lower panel)}{observable $\avg{Ib}_\text{line}^2$ (middle panel, same legend as upper panel) and lensing and IA kernels for LSST Y1- and Y10-like galaxy distributions (lower panel) all as functions of redshift}.}
    \label{fig:observables}
\end{figure}

At this juncture we have defined both LIM observables and WL observables, which we illustrate in~\autoref{fig:observables}. Now we move to consider them in Fourier space.

Since cosmic shear data are either obtained in fairly coarse tomographic bins (with source redshift errors of several percent being typically expected from photometry, per the LSST DESC SRD) or non-tomographic altogether, and we will only consider the latter scenario in this work, we will deal not with the three-dimensional power spectrum $P(\mathbf{k})$ as a function of comoving wavevector $\mathbf{k}$, but with the angular power spectrum $C_\ell$ as a function of angular multipole $\ell$.

We assume a minimum accessible central angular multipole value of $\ell_\text{min}=200$ (optimistic for the FYST DSS-like survey given its small $f_\text{sky}$ and the likely loss of large-scale angular modes in mapmaking, but more than reasonable for the Stage 3 concept), and assume the maximum workable angular multipole is $\ell_\text{max}=2000$. This is not due to observational limits but rather to avoid having to model baryonic and other non-linear effects beyond the scope of this work. We take $C_\ell$ to be calculated in 24 logarithmically evenly spaced bins of $\Delta(\ln{\ell}) = \Delta\ell/\ell\approx0.105$ between those two values of $\ell$, so that the lowest $\ell$ bin approximately spans $\ell\in(190,210)$ and the highest $\ell\in(1900,2100)$.

A field with an isotropic spherically-averaged power spectrum $P(k)$ has corresponding projected angular power spectrum $C_\ell$ found by integrating in comoving distance $\chi(z)$:
\begin{equation}
    C_\ell = \int \dd\chi\,\frac{W(\chi)^2P(k=\ell/\chi,\chi)}{\chi^2},
\end{equation}
with $W(\chi)$ being a selection function describing the redshift distribution of the signal such that $\int \dd\chi\,W(\chi)=1$. This is only true in the Limber approximation~\citep{Limber53,Kaiser92,Kaiser98,LoVerdeAfshordi08}, applicable at small scales or large $\ell$. We will be dealing with values of $\ell\in(200,2000)$ and the use of the approximation is acceptable here\added{, whether $P(k)$ is an auto-correlation or cross-correlation power spectrum}.

We note a few additional simplifications at this point. First, again since $\ell\gg1$, often we will simply use $\ell$ where $\ell+1$ or $\ell+1/2$ should technically be used. In addition, although redshift-space distortions (RSD) can significantly affect the line-intensity mapping signal as previous works by~\cite{Bernal19b,Bernal19a} and~\cite{SchaanWhite21} have noted, we assume that the line-of-sight resolution in the LIM data is sufficiently coarse for these effects to be subdominant. Work by~\cite{Jalilvand20} mostly supports this assumption, with some effect from RSD discernible for redshift bins of $\Delta z=0.01$ but only exceeding factor-of-order-unity enhancements for $\Delta z=0.001$ (which is much narrower than the width per channel for most of our CO lines for most of the 200--300 GHz observing frequency range).

Another simplification we will make is in ignoring the angular beam of the LIM instrument. Even though we work in large $\ell$, we do not work with such large $\ell$ that the effect of the beam becomes significant. Were it to be significant, and assuming the cosmic shear field has not been filtered to somehow match the LIM beam size, the observed LIM--shear cross-correlation angular power spectrum would be attenuated through multiplication by the factor
\begin{equation}
    B_\ell=\exp{\left[-\theta_\text{FWHM}^2\ell(\ell+1)/(16\ln{2})\right]},
\end{equation}
with $\theta_\text{FWHM}$ being the full width at half maximum of the angular beam. The LIM auto-correlation signal $C_\ell$ would in turn be multiplied by $B_\ell^2$. However, the~\cite{FYST} specifications suggest an expected range of $\theta_\text{FWHM}=0.7$--$1.1$ arcmin across the 200--300 GHz frequency range, and across the range of $\ell$ that we work in, we find that $B_\ell\approx1$ to within a few percent. Therefore we will ignore the effect of the LIM instrument beam profile for the remainder of this work.

Finally, in calculating this projection for LIM observables in each channel, we assume a flat selection function. The width $\Delta\chi$ of the shell in comoving space carved out by a channel with central frequency $\nu_\text{obs}$ is given by
\begin{equation}
    \Delta\chi = \frac{c(1+z)}{H(z)}\frac{\delta\nu}{\nu_\text{obs}}.
\end{equation}
Then $W(\chi)=1/\Delta\chi$ within $\pm\Delta\chi/2$ of the central $\chi$ for each channel and zero elsewhere.\added{ As an example, in the CO(4--3) observing frame, one would obtain $\chi\approx3000$ comoving Mpc and $\Delta\chi\approx50$ comoving Mpc around the middle of the frequency range.}

There are two sets of observed angular power spectra to consider at each observing frequency $\nu_\text{obs}$---the LIM auto $C_\ell$ and the LIM--WL cross $C_\ell$. In addition, there is a single (non-tomographic) WL auto $C_\ell$, and for covariance purposes we will also need to obtain expected cross spectra between different frequency channels of the LIM observation. The remainder of this section outlines in succession the calculation of each of these sets of $C_\ell$.

Note, however, that the main text of this work will not consider 1-halo terms---the clustering of galaxies and thus of line emission within an individual halo, rather than 2-halo terms describing halo--halo pairwise correlations. \cite{SchaanWhite21} do consider 1-halo terms throughout their work but do not appear to find an appreciable 1-halo component in the CO power spectrum at $z\lesssim2$, and our own explicit calculations in~\hyperref[sec:appendix-1h]{Appendix~\ref{sec:appendix-1h}} confirm that the inclusion of 1-halo terms do not significantly affect the results of this work.

\subsubsection{LIM auto spectra}
The observed LIM auto spectrum in each frequency channel, with central frequency $\nu_\text{obs}$, is given by the sum of the noise spectrum and the signal $C_\ell$ from each observed spectral line:
\begin{align}
    C_{\ell,\text{LIM}}(\nu_\text{obs}) &= C_{\ell,\text{\cii}}(\nu_\text{obs}) + \sum_{J=3}^7 C_{\ell,\text{CO},J}(\nu_\text{obs})+ C_{\ell,\text{noise}},
\end{align}
where again in the interest of brevity we are including the \ci{} $C_\ell$ as part of the CO lines, i.e., effectively the sum is over $J=\{3,4,4.27,5,6,7\}$.

The individual signal $C_\ell$ values are calculated readily from the integrated quantities previously described. Given the previously assumed flat per-channel selection function,
\begin{align}
    C_{\ell,\text{\cii}} = \frac{\avg{Ib}_\text{\cii}^2P_\mathrm{m}(k=\ell/\chi,\chi)+P_\text{shot,\cii}}{\chi^2\,\Delta\chi},
\end{align}
with $\avg{Ib}$, $\chi$, $\Delta\chi$, $P_\mathrm{m}$ (the linear matter power spectrum), and $P_\text{shot}$ all obtained at $z=\nu_\text{rest,\cii}/\nu_\text{obs}-1$. And in just the same way,
\begin{align}
    C_{\ell,\text{CO},J} = \frac{\avg{Ib}_{\text{CO},J}^2P_\mathrm{m}(k=\ell/\chi,\chi)+P_{\text{shot,CO},J}}{\chi^2\,\Delta\chi},
\end{align}
again with all comprising quantities obtained at the appropriate redshift of $z=\nu_{\text{rest,CO}(J\to J-1)}/\nu_\text{obs}-1$.

As for the noise power spectrum, this 
\begin{equation}
    C_{\ell,\text{noise}} = \frac{4\pi f_\text{sky}\sigma_\text{spec}^2}{N_\text{spec,eff}t_\text{surv}},
\end{equation}
with $\sigma_\text{spec}$ defined previously in~\autoref{sec:expcontext} and the other parameters also laid out in~\autoref{tab:expcontext}.
\subsubsection{LIM--WL cross spectra}
For the LIM--WL cross spectra we assume that the change in $W_\kappa$ and $W_\text{IA}$ within each channel is negligible across the $\Delta z\sim10^{-2}$ channel width for the CO lines. So for a channel with central observing frequency $\nu$, we gather from both~\cite{SchaanWhite21} and~\cite{Shirasaki21} that
\begin{align}
    C_{\ell,\text{LIM}\times\text{WL}}(\nu) &= \sum_{J=3}^7\frac{\avg{Ib}_{\text{CO},J}b_\mathrm{m}[W_\kappa(\chi)+W_\text{IA}(\chi)]P_\mathrm{m}\left(k=\frac{\ell}{\chi}\right)}{\chi^2}\nonumber\\&\quad+\frac{\avg{Ib}_{\text{\cii{}}}b_\mathrm{m}[W_\kappa(\chi)+W_\text{IA}(\chi)]P_\mathrm{m}\left(k=\frac{\ell}{\chi}\right)}{\chi^2},\label{eq:Ccross}
\end{align}
where given $J$ and $\nu$, each term in the sum\added{ (including the matter power spectrum)} is calculated at $z=\nu_{\text{rest,CO}(J\to J-1)}/\nu-1$ and the \cii{} term is similarly evaluated at $z=\nu_\text{rest,\cii{}}/\nu-1$. In addition, $b_m$ is the mass-averaged halo bias:
\begin{equation}
    b_\mathrm{m} = \frac{\int \dd{M_h}\,(\dd{n}/\dd{M_h})\,M_h\,b(M_h)}{\int \dd{M_h}\,(\dd{n}/\dd{M_h})\,M_h}.\label{eq:matterbias}
\end{equation}

Per~\cite{SchaanWhite21} (but ignoring the halo density profile and redshift-space effects), in principle it should be the case that given the cosmic mean matter density $\bar\rho_\mathrm{m}$, $\int \dd{M_h}\,(\dd{n}/\dd{M_h})\,M_h/\bar\rho_\mathrm{m}=1$ and thus $b_\mathrm{m}=\int \dd{M_h}\,(\dd{n}/\dd{M_h})\,(M_h/\bar\rho_\mathrm{m})\,b(M_h)$. However, as~\cite{SchaanWhite21} themselves state in Footnote 1, in most cases it is unrealistic to expect $\int \dd{M_h}\,(\dd{n}/\dd{M_h})\,M_h/\bar\rho_\mathrm{m}=1$ for most $\dd{n}/\dd{M_h}$ models and integration limits. We have therefore mirrored a correction that~\cite{SchaanWhite21} apply, albeit to a different quantity. Ultimately we find $b_\mathrm{m}\approx1$ to within a few percent, which is perhaps not entirely surprising, but we calculate it explicitly nonetheless.

Note that the \cii{} contribution to $C_{\ell,\text{LIM}\times\text{WL}}$ is negligible in practice (although we do calculate it explicitly throughout this work) simply because at \cii{} redshifts, $\chi$ is much larger and the WL kernel amplitude much smaller.
\subsubsection{WL auto spectrum}
Unlike the LIM auto and LIM--lensing cross spectra, the lensing auto spectrum is from an entirely projected map, so we integrate across all $\chi$ rather than evaluating quantities at specific values of $\chi$. As indicated by~\cite{Shirasaki21}, the total observed power spectrum is
\begin{align}
    C_{\ell,\text{WL}}&=\int \dd\chi\,\frac{[W_\kappa(\chi)+W_\text{IA}(\chi)]^2}{\chi^2}P_\text{NL}(k=\ell/\chi,\chi)+\frac{\sigma_e^2}{n_g}.
\end{align}
Note the use of the non-linear matter power spectrum $P_\text{NL}$, which \texttt{hmf} calculates using \texttt{HALOFIT}~\citep{HALOFIT}, in lieu of the linear $P_m$ used for our halo models. In practical terms, we could not use the non-linear power spectrum with the LIM observables in our present formalism given the use of linear bias throughout. From a forecasting perspective, this is a conservative choice as we choose to not model any non-linear component to the cross-correlation signal between line-intensity data and either cosmic shear or IA, and thus potentially reduce the forecast signal relative to covariances.

\subsubsection{LIM--LIM cross spectra}
For the cross spectra between different channels of the LIM data, we assume that all noise is uncorrelated between frequencies and the only source of cross-frequency correlation is in CO lines mapped in overlapping volumes. The validity of this simplification will depend on the nature of atmospheric and instrumental fluctuations, as well as the ability of analysis pipelines to remove correlated noise and systematics. But assuming all these experimental aspects are favourable, following~\cite{Shirasaki21} we have
\begin{align}
    C_{\ell,\text{LIM}\times\text{LIM}}(\nu,\nu') &= \delta_{\nu\nu'}C_{\ell,\text{noise}}+\delta_{\nu\nu'}C_{\ell,\text{\cii{}}}(\nu)+{}\nonumber\\&\sum_{J=3}^7\sum_{J'=3}^7\Theta{(\nu,\nu',J,J')}\times\nonumber\\&\quad\frac{\avg{Ib}_{\text{CO},J}\avg{Ib}_{\text{CO},J'}P_\mathrm{m}\left(k=\frac{\ell}{\bar\chi}\right)+P_{\text{shot},J,J'}}{\bar\chi^2\,\overline{\Delta\chi}},
\end{align}
where we have introduced a significant amount of new notation that will need to be unpacked bit by bit. First, $\Theta$ selects for channels where a majority of the sampled redshift interval in one CO line (or the \ci{} line) overlaps with that in another line:
\begin{equation}
    \Theta{(\nu,\nu',J,J')} = \begin{cases}1&\text{if }\left|\frac{J}{\nu}-\frac{J'}{\nu'}\right|<\frac{\delta\nu}{2\nu}\sqrt{\frac{JJ'}{\nu\nu'}},\\0&\text{otherwise.}\end{cases}
\end{equation}
(We have simplified the condition statement by dividing out $\nu_\text{rest,CO(1--0)}$ from both sides of the inequality.) Since each term in the sum is only nonzero if the respective CO lines in the respective pair of channels originate from the same redshift (within channel bandwidth), each term is thus calculated at that redshift. The central values of $\chi$ and $\chi'$ at frequencies $\nu$ for CO($J\to J-1$) and $\nu'$ for CO($J'\to J'-1$) may be slightly different, and the same may be true for the shell widths $\Delta\chi$ and $\Delta\chi'$. So we have used the notations $\bar\chi$ and $\overline{\Delta\chi}$ to denote the geometric mean between the two different values. Note that because we set $\delta\nu/\nu=100$ as a constant, no such averaging is done for that quantity in the inequality defining $\Theta$.

Finally, the cross shot noise $P_{\text{shot},J,J'}$ between the two CO lines is given similarly to the auto shot noise:
\begin{align}
    P_{\text{shot},J,J'}&=C_\text{LI,CO}(\nu,J)C_\text{LI,{CO}}(\nu',J')\exp{\left(\sigma_{\text{CO},J}\sigma_{\text{CO},J'}\ln^2{10}\right)}\nonumber\\&\qquad\times\int \dd{M_h}\,\frac{\dd{n}}{\dd{M_h}}\,L_{\text{CO},J}(M_h)L_{\text{CO},J'}(M_h),
\end{align}
where for brevity we define for the CO lines
\begin{align}
    C_\text{LI,{CO}}(\nu,J)\equiv C_\text{LI}(\nu_\text{rest,CO(1--0)}J,\nu_\text{rest,CO(1--0)}J/\nu-1),
\end{align}
and we have simplified an expression from~\cite{Yang21} based on assumptions we established previously that any log-normal scatter around mean $L(M_h)$ relations is perfectly correlated between lines and that $\sigma_{\text{CO},J}=0.4$ (in units of dex) for all $J$.

\section{Forecasts}
\label{sec:forecasts}

Having established relevant experimental parameters in~\autoref{sec:expcontext} and model parameters in~\autoref{sec:models}, we can now calculate the expected signal and covariance, as well as consider possible inferences.
\subsection{Signal and detectability}
Per~\cite{Shirasaki21}, the covariance matrix for LIM--WL cross spectra in multipoles $\ell$ and $\ell'$ across frequencies $\nu$ and $\nu'$ is given by
\begin{equation}\mathbf{C}_{\ell\ell'}(\nu,\nu')=\delta_{\ell\ell'}\cdot\frac{C_{\ell,\text{LIM}\times\text{WL}}^2(\nu)+C_{\ell,\text{LIM}\times\text{LIM}}(\nu,\nu')C_{\ell,\text{WL}}}{2\ell\,\Delta\ell f_\text{sky}}.\end{equation}
All $C_\ell$ here are total observed spectra, including both signal and noise contributions as we have done previously. This \added{covariance }will always be minimised through larger $f_\text{sky}$, as the only part of the numerator that scales with $f_\text{sky}$ is $C_{\ell,\text{noise}}\propto f_\text{sky}$ as part of the sum in~\autoref{eq:Ccross} that defines $C_{\ell,\text{LIM}\times\text{LIM}}(\nu,\nu')$. So although we chose $f_\text{sky}$ conservatively in~\autoref{sec:expcontext} for the mm-wave Stage 3 survey, increasing it all the way up to $\sim10^{-1}$ to match the LSST sky coverage would non-negligibly boost the cross $C_\ell$ detection significance.

\added{For the purposes of this paper, we neglect non-Gaussian covariance, including super-sample covariance (i.e., contributions to covariance from super-survey modes---cf.~\citealt{TakadaHu13,Barreira18}). We suppose such non-Gaussian terms have at most order-of-unity effects on the conclusions of this paper, and leave to future work a detailed examination of these terms and their relevance to LIM--WL cross-correlation.}

Defining the covariance matrix allows us to evaluate the total detection significance as
\begin{equation}
    \left(\frac{\mathrm{S}}{\mathrm{N}}\right)_\text{total} = \sqrt{\sum_{\ell,\ell',\nu,\nu'} C_{\ell,\text{LIM}\times\text{WL}}(\nu)\mathbf{C}^{-1}_{\ell\ell'}(\nu,\nu')C_{\ell',\text{LIM}\times\text{WL}}(\nu')}.
\end{equation}

\added{\begin{figure}
    \centering
    \includegraphics[clip=True,trim=2mm 5mm 5mm 2mm,width=0.96\linewidth]{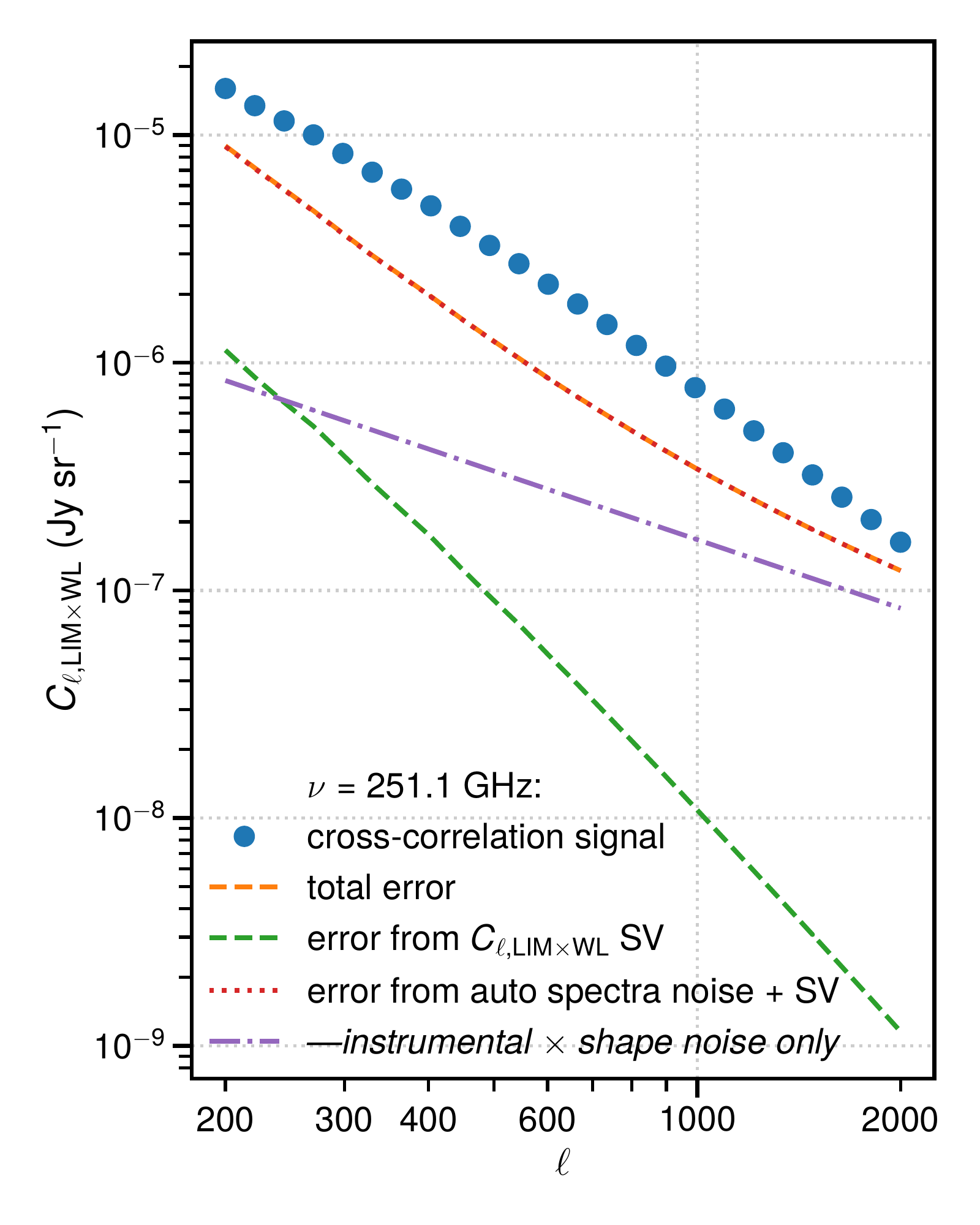}
    \caption{The LSST Y10-like WL $\times$ mm-wave Stage 3 LIM cross-correlation signal $C_{\ell,\text{LIM}\times\text{WL}}(\nu)$ at $\nu=251$ GHz (large dots), shown alongside the total error (orange dashed), the error from the cross signal SV alone (green dashed), the error from the total observed auto-correlation signals (including both SV and noise contributions; dotted), and the portion of the error budget attributable purely to the combination of LIM instrumental noise and WL shape noise (dashed-dotted).}
    \label{fig:sn_perch_detail}
\end{figure}

We can see from the expression for $\mathbf{C}_{\ell\ell'}(\nu,\nu')$ that the error on the cross spectrum $C_{\ell,\text{LIM}\times\text{WL}}(\nu)$ in each channel, given by $\mathbf{C}_{\ell\ell'}^{1/2}(\nu,\nu)$, is a mix of sample variance (SV), for both auto and cross spectra, and the instrumental and shape noise contributions to the observed auto spectra outlined in~\autoref{sec:celloutline}. \autoref{fig:sn_perch_detail} illustrates the amplitude of each of these components relative to the cross signal and total per-channel error by breaking down the signal and error for $\nu=251$ GHz, in the context of the LSST Y10-like WL $\times$ mm-wave Stage 3 LIM scenario. }

Because the observed $C_{\ell,\text{LIM}\times\text{LIM}}(\nu,\nu')$ values include contributions from multiple bright lines and not all of them will correlate significantly with the WL data, the LIM measurement contributes significantly to covariance not only through instrumental noise but also through these non-correlating lines. \replaced{This renders}{\autoref{fig:sn_perch_detail} shows in particular how the combination of instrumental and shape noise by themselves contribute at most half of the error and even then only at high $\ell$. Such characteristics of this observation render} the LIM--WL cross spectrum \emph{more} difficult to detect than LIM auto spectra, and far from a quick path to early science output for LIM experiments.

As such, we forecast a total LIM $\times$ WL cross-correlation detection significance of 1.8 for LSST Y1-like $\times$ FYST DSS-like parameters. However, this rises sharply to 50 for LSST Y10-like $\times$ mm-wave Stage 3 parameters. While a marginal detection of the overall signal is therefore possible in principle with the current generation of instruments, we expect future LIM experiments with substantially greater sensitivity and sky coverage to enable a firm high-significance detection of this cross-correlation.

\begin{figure}
    \centering
    \includegraphics[clip=True,trim=2mm 5mm 5mm 2mm,width=0.96\linewidth]{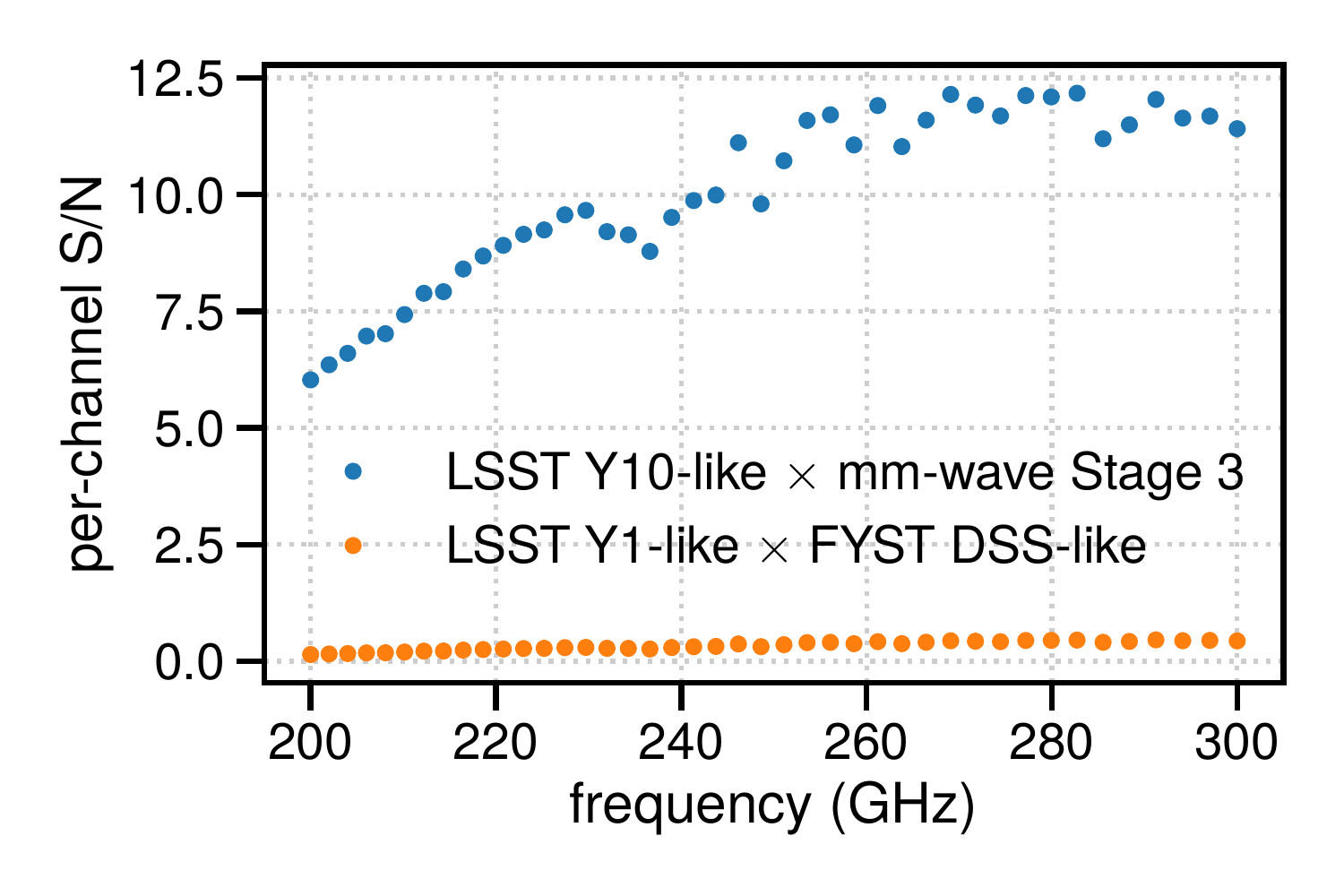}
    \caption{Per-channel detection significance of $C_{\ell,\text{LIM}\times\text{WL}}(\nu)$, calculated for both experimental scenarios as indicated in the legend.}
    \label{fig:sn_perch}
\end{figure}

We also show in~\autoref{fig:sn_perch} the signal-to-noise ratio per channel, calculated as
\begin{equation}
    \left(\frac{\mathrm{S}}{\mathrm{N}}\right)(\nu) = \sqrt{\sum_{\ell,\ell'} C_{\ell,\text{LIM}\times\text{WL}}(\nu)\mathbf{C}^{-1}_{\ell\ell'}(\nu,\nu)C_{\ell',\text{LIM}\times\text{WL}}(\nu)}.
\end{equation}
Note that even the detection significance in each channel exceeds $5\sigma$ in the LSST Y10-like $\times$ mm-wave Stage 3 scenario.

\begin{figure*}
    \centering
    \includegraphics[width=0.96\linewidth]{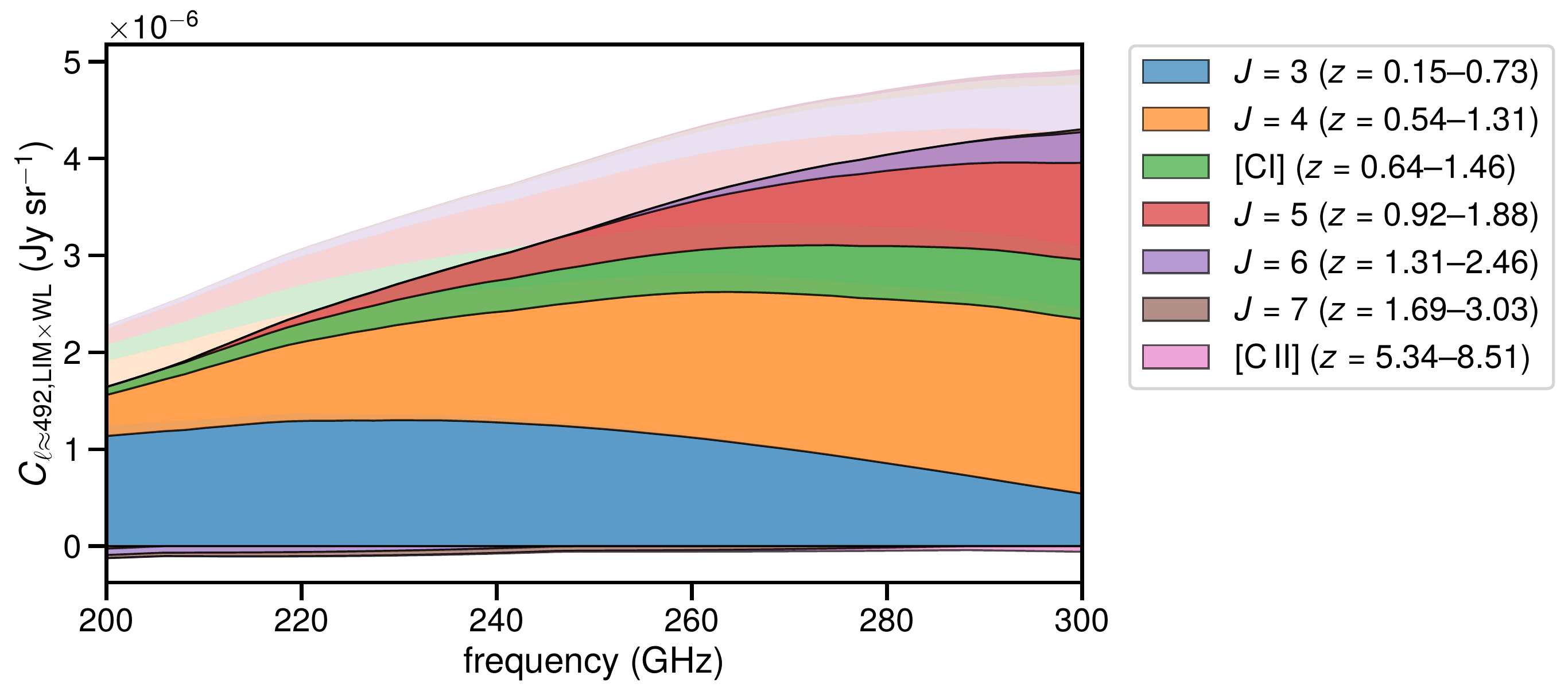}
    \caption{The LIM--WL cross-correlation $C_\ell$ at a central value of $\ell\approx492$, shown as a function of observing frequency and broken down into contributions from individual lines (shaded areas with black edges, as indicated in the legend). Values below zero represent negative contributions to the overall signal. We also show the signal prediction when intrinsic alignments are unaccounted for (shown in fainter shaded areas without black edges). For this figure, we assume a LSST Y10-like source distribution to derive WL observables.}
    \label{fig:cell_break}
\end{figure*}

Given the high signal-to-noise eventually expected, it is interesting to examine the signal in a little more detail. To illustrate the contribution of different CO lines to the total cross-correlation signal, we consider $C_{\ell,\text{LIM}\times\text{WL}}(\nu)$ at $\ell\approx492$ as would be detected using a LSST Y10-like source distribution, but broken down into the individual terms of the summation in~\autoref{eq:Ccross}.\footnote{\added{The choice of $\ell\approx492$ is arbitrary and does not affect the qualitative characteristics of this illustration.}} We show a stacked plot of these terms in~\autoref{fig:cell_break}. We see that there is truly tomographic astrophysical cross-correlation at work, with each channel probing the contribution of different CO lines from different redshifts. At higher frequencies the CO(3--2) line from $z\sim0.2$ is subdominant to higher-$J$ lines from $z\sim0.5$--1, while at lower frequencies CO(3--2) from $z\sim0.6$ and CO(4--3) at $z\sim1$ comprise the bulk of the signal.

We also show in~\autoref{fig:cell_break} what the signal would have looked like without the contribution of IA terms, which is negative because of the orthogonal direction of IA and shear-derived correlations as noted in~\autoref{sec:IAmodel}. As the relative amplitudes of the WL and IA kernels shown in~\autoref{fig:observables} would suggest, the impact is not insignificant for cross-correlation at $z\gtrsim1$ where $W_\kappa$ falls with increasing redshift while $-W_\text{IA}$ is still rising with redshift.

In all scenarios, the CO(7--6) and \cii{} lines essentially do not contribute to the signal. We expect this to be the case for both lines, due to a combination of low brightness and largely mismatched source galaxy distributions. The latter also explains why the CO(5--4) and CO(6--5) lines, despite being two of the three brightest lines across the observing band as shown in~\autoref{fig:observables}, contribute relatively little to the cross $C_\ell$ with or without intrinsic alignments. This echoes our point earlier about significant covariance contributions from observed lines that are bright but uncorrelated with WL data.

\subsection{Inferences}
\label{sec:mcmc}
As shown above, the cross-correlation signal is the outcome of several different astrophysical lines tracing large-scale structure across a range of redshifts with time-varying amplitude. There are indeed many parameters we could attempt to constrain with these data, but given the highly preliminary nature of these forecasts we will consider inference of a fairly limited set of parameters for the time being. In particular, we consider the constraining power of the $C_{\ell,\text{LIM}\times\text{WL}}$ measurement with respect to three parameters (fixing all other model parameters at fiducial values): the $L_\text{IR}(L_\text{CO})$ power-law parameters $\alpha_\text{IR--CO}$ and $\beta_\text{IR--CO}$, and the IA effect amplitude $A_\text{IA}$.

A Fisher forecast, which we detail in~\hyperref[sec:appendix-fisher]{Appendix~\ref{sec:appendix-fisher}}, establishes that only the mm-wave Stage 3 survey is sensitive enough to constrain parameters beyond priors from a cross-correlation. To characterise the posterior distribution in that scenario, we run a Markov chain Monte Carlo (MCMC) calculation using \texttt{emcee}~\citep{emcee,emceev3}, a Python implementation of the affine-invariant ensemble sampling algorithm of~\cite{GW10}.

We impose a flat prior of $\alpha_\text{IR--CO}\in[0.5,2]$ following~\cite{Sun21} but do not impose a prior on $\beta_\text{IR--CO}$ as we expect a strong degeneracy between $\alpha_\text{IR--CO}$ and $\beta_\text{IR--CO}$ based on the Fisher forecast of~\hyperref[sec:appendix-fisher]{Appendix~\ref{sec:appendix-fisher}}. We also impose a Gaussian prior of $A_\text{IA}=1.0\pm1.0$, meant to be broad and weakly suggest weak evidence for a positive value of order unity without being overly informative.

Our MCMC with \texttt{emcee} uses 32 walkers to navigate the space of parameters $\{\lambda_i\}=\{\alpha_\text{IR--CO},\beta_\text{IR--CO},A_\text{IA}\}$. The log-probability function (up to an additive constant) is taken to be $-\infty$ if $\alpha_\text{IR--CO}<0.5$ or $\alpha_\text{IR--CO}>2$, and otherwise
\begin{align}
    &\ln{p(\lambda_i)} = -\frac{1}{2}(A_\text{IA}-1)^2\nonumber\\&\ {}-\frac{1}{2}\sum_{\ell,\ell',\nu,\nu'} \Delta C_{\ell,\text{LIM}\times\text{WL}}(\nu|\lambda_i)\mathbf{C}^{-1}_{\ell\ell'}(\nu,\nu')\Delta C_{\ell',\text{LIM}\times\text{WL}}(\nu'|\lambda_i),
\end{align}
where $\Delta C_{\ell,\text{LIM}\times\text{WL}}(\nu|\lambda_i)$ is the difference between the `measured' $C_{\ell,\text{LIM}\times\text{WL}}(\nu)$ (here taken simply to be the fiducial spectrum) and the predicted values based on proposed parameter values $\lambda_i$.

We run the MCMC sampler for 10000 steps. Based on walker-averaged autocorrelation times of 130--160 steps for each parameter, we conservatively discard the first 3200 steps and consider the remaining 6800 steps to adequately sample the posterior distribution in a converged manner.

\begin{figure}
    \centering
    \includegraphics[width=0.96\linewidth]{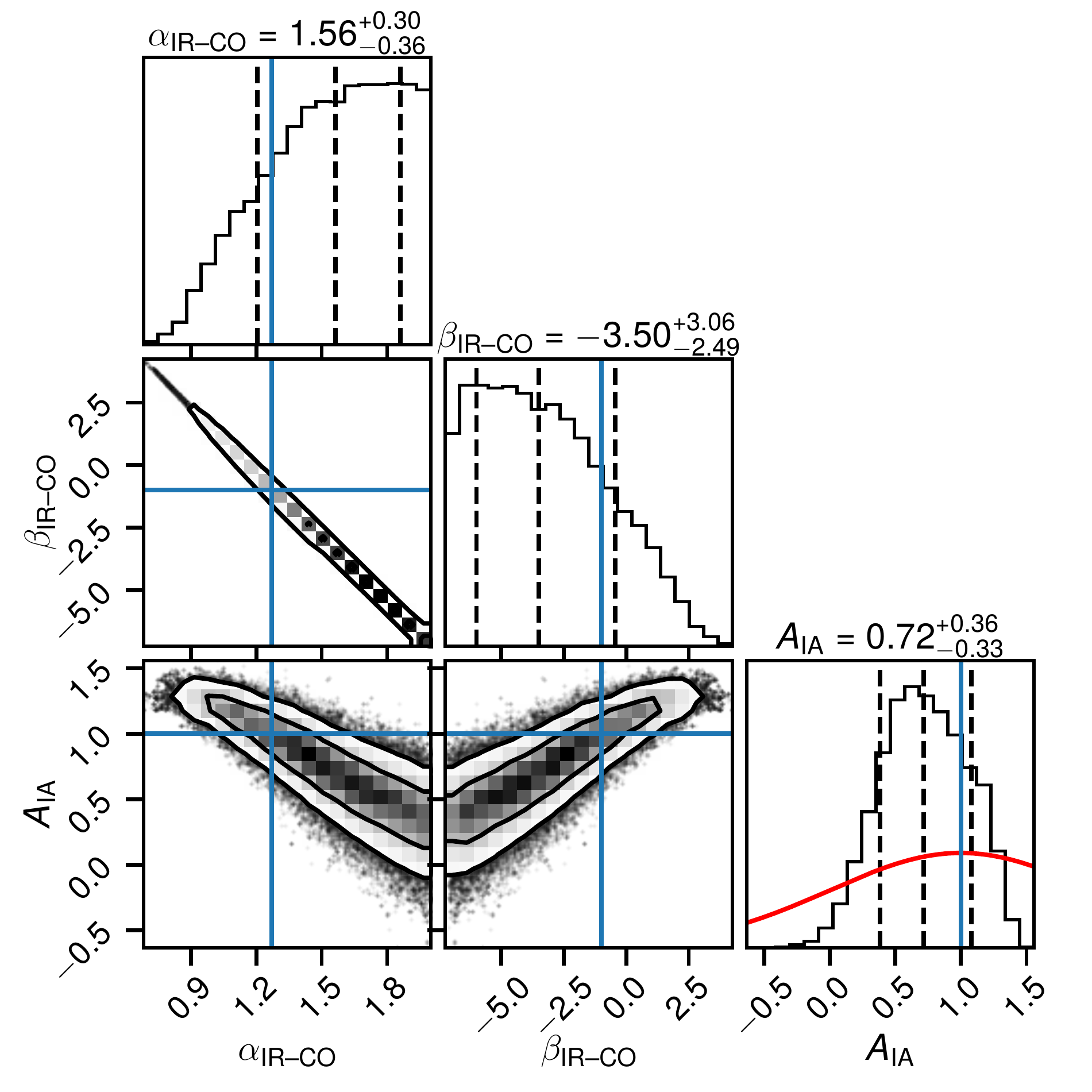}
    \caption{68\% (inner contours) and 95\% (outer contours) confidence regions for the model parameters as constrained by the LSST Y10-like WL $\times$ mm-wave Stage 3 LIM cross-correlation signal. We also indicate marginalised 68\% confidence intervals around the median estimate for each parameter (vertical dashed lines), fiducial model values (blue solid lines), and the broad Gaussian prior on $A_\text{IA}$ (red curve in bottom rightmost panel).}
    \label{fig:emcee_corner}
\end{figure}

We show the resulting posterior distribution estimate in~\autoref{fig:emcee_corner}. The credible regions for the astrophysical parameters $\alpha_\text{IR--CO}$ and $\beta_\text{IR--CO}$ are highly skewed, and as such the constraining power on them is modest and truncated by our flat prior for $\alpha_\text{IR--CO}$ on one side. Our constraints still place a meaningful one-sided limit on the $L_\text{CO}(L_\text{IR})$ power-law slope, and would likely be tighter were we to also consider the LIM auto signal and its constraining power---taking advantage of line-of-sight modes---either in isolation or joint with the LIM--WL cross signal.

Perhaps more interesting, however, are the predicted constraints related to $A_\text{IA}$. First, with LSST Y10-like WL $\times$ mm-wave Stage 3 LIM data, we would constrain $A_\text{IA}$ well beyond the prior width, with a marginalised 68\% confidence interval of $A_\text{IA}=0.72^{+0.36}_{-0.33}$ consistent with the fiducial value of 1. But beyond just looking at $A_\text{IA}$ by itself, we find a \emph{strong degeneracy between the astrophysical parameters and $A_\text{IA}$}. This is a sensible result when reflecting on the effect of IA as shown in~\autoref{fig:cell_break}---a lower cross-correlation signal could easily be due to either a dimmer CO signal (i.e., higher $\alpha_\text{IR--CO}$ values\footnote{In principle, a similar degeneracy should exist between $\beta_\text{IR--CO}$ and $A_\text{IA}$ but our calculations suggest that any such degeneracy is insignificant next to the strong degeneracy between $\alpha_\text{IR--CO}$ and $\beta_\text{IR--CO}$.}) or a stronger IA effect (i.e., higher $A_\text{IA}$ values). Therefore, if external galaxy survey data can better characterise the IA effect, we would have stronger constraints on astrophysics. Conversely, more information on astrophysics from LIM experiments or other surveys would actually allow LIM--WL cross-correlation to strongly constrain the IA model, although this is admittedly more speculative and depends on other cosmological parameters being constrained tightly\added{. An exploratory Fisher forecast considered at the end of~\hyperref[sec:appendix-fisher]{Appendix~\ref{sec:appendix-fisher}} suggests that including astrophysical information from the mm-wave Stage 3 LIM auto spectra could improve the IA amplitude constraint to a level of $\sigma[A_\text{IA}]<0.1$, a highly interesting proposition that future work will hopefully verify while accounting properly for covariances between the various signals being measured}.

\begin{figure*}
    \centering
    \includegraphics[width=0.96\linewidth]{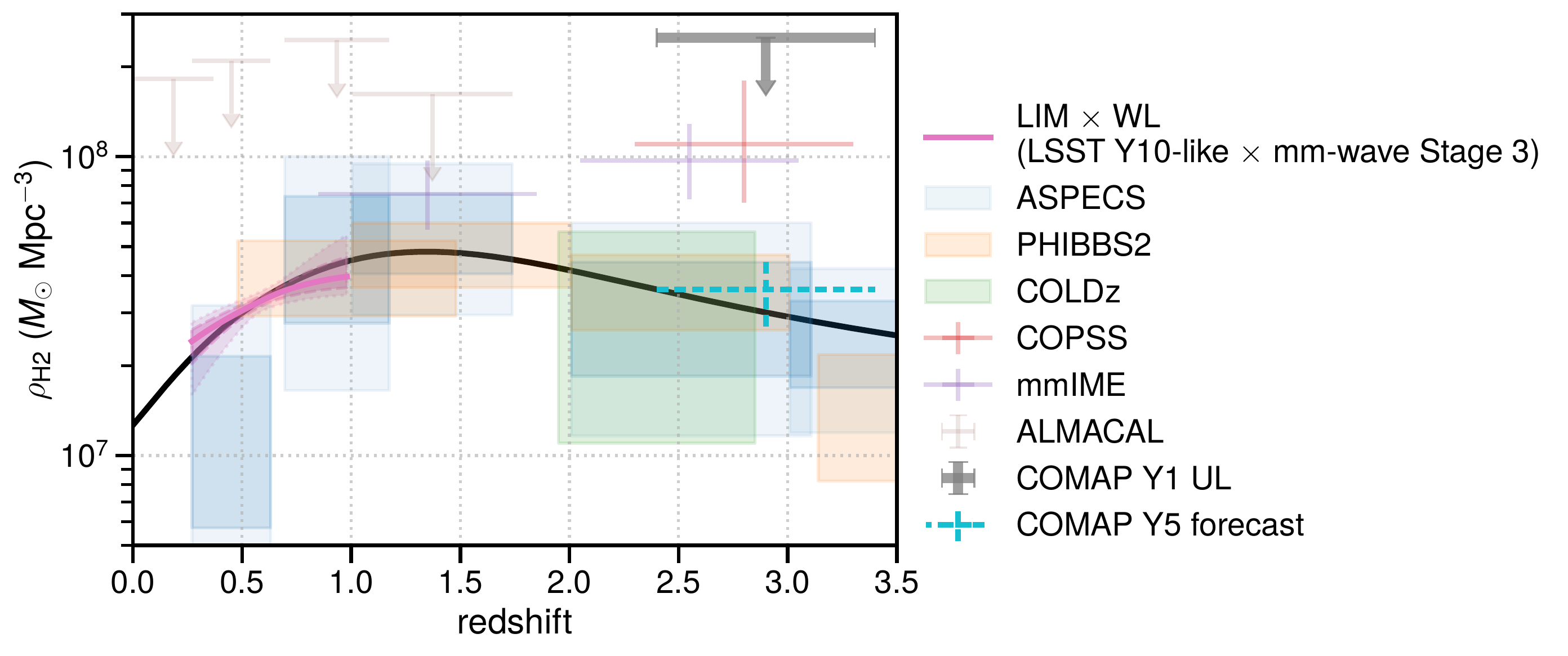}
    \caption{Approximate constraints expected from LIM--WL cross-correlation on $\rho_\text{H2}$, shown as 68\% and 95\% credible intervals (magenta shaded areas bounded by dashed and dotted curves, respectively) around the median $\rho_\text{H2}$ value (magenta solid curve). We compare these constraints to both the fiducial model expectation (black solid curve) and observational results from ASPECS~\citep{ASPECS-LPLF2}, PHIBBS2~\citep{PHIBBS2Lenkic}, COLDz~\citep{COLDzLF}, COMAP~\citep{Cleary21}, COPSS~\citep{COPSS}, mmIME~\citep{mmIME-ACA}, and ALMACAL~\citep{Klitsch19}. All results use $\alpha_\text{CO}=3.6\,M_\odot\,($K\,km\,s$^{-1}$\,pc$^{-2})^{-1}$ except COPSS and mmIME, which use a conversion of $\alpha_\text{CO}=4.3\,M_\odot\,($K\,km\,s$^{-1}$\,pc$^{-2})^{-1}$.}
    \label{fig:rhoH2_illust}
\end{figure*}

\begin{figure}
    \centering
    \includegraphics[width=0.96\linewidth]{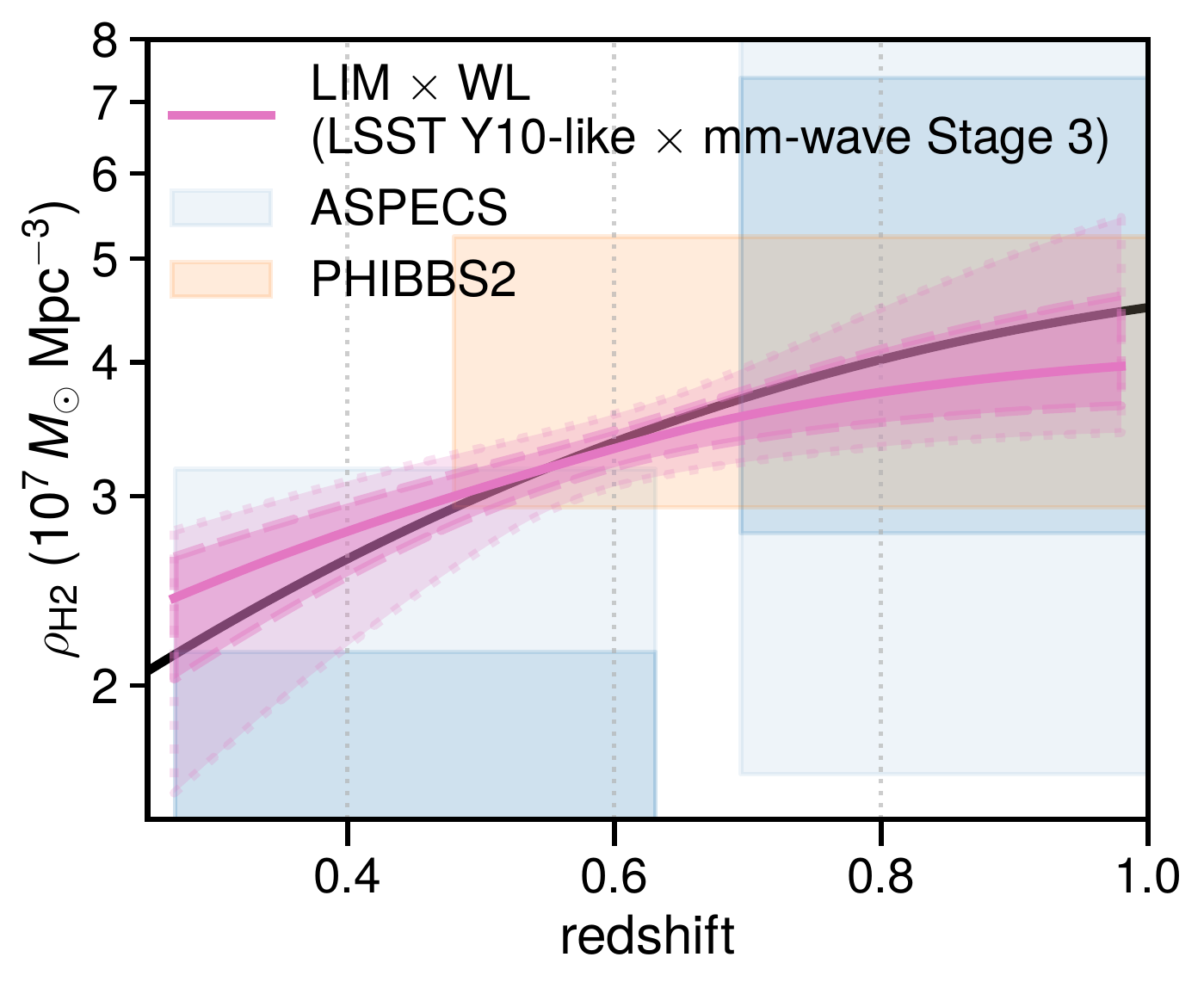}
    \caption{Same as~\autoref{fig:rhoH2_illust}, but across a narrower range of redshifts ($z=0.25$--1.0) and $\rho_\text{H2}$ to more clearly show constraints expected from LIM--WL cross-correlation.}
    \label{fig:rhoH2_illust_detail}
\end{figure}

As the forecast stands, we can approximately illustrate the constraining power forecast in~\autoref{fig:emcee_corner} on the IR--CO power-law relation in more concrete terms, by calculating what the range of predictions would be for the cosmic H$_2$ mass density $\rho_\text{H2}$ given the posterior samples obtained for the CO--IR luminosity relation. Assuming a conversion factor $\alpha_\text{CO}=3.6\,M_\odot\,($K\,km\,s$^{-1}$\,pc$^2)^{-1}$ \added{(following, e.g.,~\citealt{COLDzLF}, who cite~\citealt{Daddi10} as the source of this value) }and integrating modelled CO(1--0) line luminosities (in units of velocity- and area-integrated brightness temperature) across the halo mass function,
\begin{equation}
    \rho_\text{H2} = \alpha_\text{CO}\int \dd{M_h}\,\frac{\dd{n}}{\dd{M_h}}\,{L'_\text{CO(1--0)}(M_h,z)},
\end{equation}
where the line luminosities are calculated based on~\autoref{eq:LCO10} and thus vary based on $\alpha_\text{IR--CO}$ and $\beta_\text{IR--CO}$. We illustrate the resulting level of constraining power on $\rho_\text{H2}$ in~\autoref{fig:rhoH2_illust}, for $z=0.27$--0.98.\footnote{These are the redshifts represented most prominently by the LSST Y10-like source distribution behind the WL signal, specifically where $n(z)$ is above 75\% of the maximum $n(z)$ value. Considering the actual WL kernel $W_\kappa(\chi)$ instead results in $z=0.23$--0.98, not a significantly different redshift range.}

The forecast suggests strong $\rho_\text{H2}$ constraints from the LIM--WL cross-correlation signal. As~\autoref{fig:cell_break} shows, the most prominently represented CO lines in this signal are CO(3--2) and CO(4--3), which are observed emitting from $z\sim0.38$ and $z\sim0.84$. At those redshifts, we predict that the cross-correlation obtains $\rho_\text{H2}(z=0.38)=2.7_{-0.3}^{+0.2}\times10^7\,M_\odot$ Mpc$^{-3}$ and $\rho_\text{H2}(z=0.84)=3.8_{-0.3}^{+0.4}\times10^7\,M_\odot$ Mpc$^{-3}$ (68\% intervals), each consistent with the corresponding fiducial model values of $2.5\times10^7\,M_\odot$ Mpc$^{-3}$ and $4.1\times10^7\,M_\odot$ Mpc$^{-3}$. This level of constraining power, which we show in more detail in~\autoref{fig:rhoH2_illust_detail}, is competitive as a notable additional probe of $z\lesssim1$ CO, alongside results from emission/absorption line candidate searches like the ALMA SPECtroscopic Survey (ASPECS) in the Hubble Ultra-Deep Field~\citep{ASPECS-LPLF2}, the Plateau de Bure High-$z$ Blue-Sequence Survey 2~\citep[PHIBBS2;][]{PHIBBS2Lenkic}, and ALMACAL~\citep{Klitsch19}, as well as higher-redshift searches like the CO Luminosity Density at High Redshift survey~\citep[COLDz;][]{COLDz,COLDzLF}, and cm- to mm-wave LIM experiments previously mentioned in this work.

\added{We only show constraints on $\rho_\text{H2}$ derived from the constraints on our CO model, in the interest of not belabouring the point. But this should not be considered the limit for the kinds of astrophysical quantities that CO LIM can statistically probe at these redshifts. For instance, we could equally as well have presented constraints on cosmic SFR density given that observations also link CO luminosities at least locally to IR luminosity and SFR, a link in fact exploited by the line emission model laid out in~\autoref{sec:linemodels} and citations therein.}

\addedtwo{Furthermore, cross-correlation against \emph{tomographic} cosmic shear power spectra---i.e., cosmic shear statistics obtained from source galaxies in different redshift slices---would potentially further refine interpretation of the evolution of CO emissivity across cosmic time. Such doubly tomographic cross-correlations, however, will require much more careful consideration of covariances across a much larger vector of observable quantities. Furthermore, the detectability of the LIM--WL cross spectrum for each tomographic WL bin would be significantly smaller than for the cross spectrum obtained in cross-correlation against non-tomographic WL data, on account of lower source galaxy counts in each bin leading to higher shape noise, redshift slicing of WL data leading to higher covariance from non-correlating CO lines, and so on. We therefore leave consideration of doubly tomographic LIM--WL cross spectra to future work, which could also consider optimal tomographic binning schemes for WL data in the context of cross-correlation with mm-wave LIM.}

\section{Conclusions}
\label{sec:conclusions}

We have now arrived at some interesting answers for the questions we posed at the start of this paper:
\begin{itemize}
    \item \emph{What is the expected detection significance of cosmic shear--LIM cross-correlation?} Although the current generation of mm-wave LIM experiments will potentially detect a cross-correlation signal against a LSST Y1-like dataset at only a $2\sigma$ level, a future mm-wave Stage 3 LIM experiment could measure the cross-correlation signal with an overall signal-to-noise of 50, with the detection significance in \emph{each individual frequency channel} in excess of $5\sigma$.
    \item \emph{What kinds of quantities could this cross-correlation constrain, astrophysical or otherwise?} We expect modest constraining power on the CO--IR luminosity relation and on the IA effect amplitude when marginalising over or fixing other parameters. Derived constraints on integrated quantities like $\rho_\text{H2}$ are competitive with other probes of CO at $z\lesssim1$. We also observe a strong degeneracy between astrophysical parameters and IA model parameters, which could be an important lever arm in joint analyses with other astrophysical and cosmological probes for improved constraints on the $z\lesssim1$ universe.
\end{itemize}

To be sure, future work must better characterise both the expected signal and its constraining power. More advanced analyses, incorporating covariances between LIM auto spectra and LIM--WL cross spectra, or even masking $z>1$ galaxies in the manner of~\cite{Sun16} to reduce cross-correlation covariance, could enhance achievable constraining power beyond what we forecast here.

In addition, this work purposefully avoids dealing with scales of $\ell>2000$ to exclude baryonic effects and highly non-linear scales. The non-linear regime could potentially provide incredibly compelling aspects to cross-correlations with the LIM signal, which in the case of star-formation lines arises from baryonic physics in dense environments, and whose cross-correlation with discrete galaxy samples would reveal average line luminosities through the cross shot noise~\citep{BreysseAlexandroff19}. However, non-linear effects undoubtedly complicate observables in highly model-dependent ways---especially in combination with RSD, which~\cite{Jalilvand20} found to partially dampen non-linear effects on $C_\ell$. Therefore, future work must provide improved, flexible modelling of both LIM and cosmic shear observables at high $k$ or $\ell$ and examine the full potential of this cross-correlation at small scales.

This work is also preliminary in that we do not examine all of the LSST data that could be cross-correlated against LIM data. Photometric galaxy surveys now commonly make use of a $3\times2$ pt analysis, leveraging two-point correlations in (tomographic) cosmic shear, galaxy--galaxy lensing, and galaxy clustering. Looking beyond the non-tomographic cosmic shear considered in this paper, LIM cross-correlations with this full $3\times2$ pt suite could have significant potential for both astrophysical and cosmological inferences, \replaced{which we leave for future work for examine}{with previous works having already developed spherical harmonic tomography techniques for LIM contexts~\citep[cf.][]{Anderson22}. We leave detailed exploration of this potential for future work to examine}.

\section*{Acknowledgements}

Thanks to Ren\'{e}e Hlo\v{z}ek for a conversation that led to the writing of this paper. Thanks also to Eric Switzer and Hamsa Padmanabhan for discussions that informed certain sensitivity scalings presented in this work. The author also gratefully acknowledges Kirit Karkare and Patrick Breysse for discussions and comments relevant to \hyperref[sec:conclusions]{the Conclusions}\replaced{, and}{;} Hamsa Padmanabhan again for a close reading of and useful comments on an earlier draft of this work\added{; Eric Switzer again for comments on the preprint version of this paper; and Marco Viero, Anthony Pullen, and Abhishek Maniyar for interesting discussions while the paper was in review}.

DTC is supported by a CITA/Dunlap Institute postdoctoral fellowship. The Dunlap Institute is funded through an endowment established by the David Dunlap family and the University of Toronto. The University of Toronto operates on the traditional land of the Huron-Wendat, the Seneca, and most recently, the Mississaugas of the Credit River; DTC is grateful to have the opportunity to work on this land. This research made use of Astropy,\footnote{http://www.astropy.org} a community-developed core Python package for astronomy \citep{astropy:2013, astropy:2018}. This research also made use of NASA's Astrophysics Data System Bibliographic Services. \added{Finally, the author would like to thank an anonymous referee for a prompt yet thoughtful review with constructive comments that helped improve this manuscript.}

\section*{Data Availability}

The code and calculations underlying this article are contained in a Jupyter notebook~\citep{jupyternb} available through a GitHub repository maintained by the author (\url{https://github.com/dongwooc/lim-lensing-predictions}). Should the GitHub repository ever become unavailable, the same code and calculations will be shared on reasonable request to the author.



\bibliographystyle{mnras}
\bibliography{ms} 




\appendix

\section{The effect of including 1-halo terms and the \replacedtwo{NFW}{halo density} profile, with respect to the present work}
\label{sec:appendix-1h}
1-halo terms in LIM auto and cross spectra arise from the clustering of line emission \emph{within} a single dark matter halo. As a result, in principle it requires assuming forms for the dark matter mass density profile and the halo mass--concentration relation. For this section, we will use the combination of \texttt{hmf} and \texttt{halomod}~\citep{hmf,halomod} to calculate the Navarro--Frenk--White\addedtwo{ (NFW)} halo density profile as modelled by~\cite{NFW} and the halo concentration as modelled by~\cite{Duffy08}, and calculate the 1-halo terms as outlined by~\cite{SchaanWhite21}.

\cite{SchaanWhite21} give the LIM--LIM cross 1-halo term in their Equation 2.11, which we adapt here to the notation of this work:
\begin{align}
    P_{\text{1h},J,J'}&=C_\text{LI,\text{CO}}(\nu,J)C_\text{LI,\text{CO}}(\nu',J')\nonumber\\&\quad\times\int \dd{M_h}\,\frac{\dd{n}}{\dd{M_h}}\,L_{\text{CO},J}(M_h)L_{\text{CO},J'}(M_h)\,|u(k,M_h)|^2,
\end{align}
where $u(k,M_h)$ is the normalised Fourier-space density profile of a halo for comoving wavenumber $k$, such that $u(k\to0,M_h)\to1$. We also neglect the redshift-space blurring of the halo profile that~\cite{SchaanWhite21} model because this will be highly subdominant in our measurement of angular power spectra in fairly low-resolution frequency channels. This expression can also be easily rewritten to describe LIM auto spectra for the CO and \ci{} lines (setting $\nu=\nu'$ and $J=J'$, recalling that we essentially treat \ci{} as if it were part of the CO line ladder with $J=4.27$) and the \cii{} line as well.

\cite{SchaanWhite21} also describe the 1-halo term for the LIM--WL cross power spectrum, which we again rewrite in our notation. Consider $C_{\ell,1h,\text{LIM}\times\text{WL}}$ to be the sum of contributions $C_{\ell,1h,\text{line}\times\text{WL}}$ corresponding to each line, where
\begin{align}
    C_{\ell,1h,\text{line}\times\text{WL}}(\nu) &= \frac{C_\text{LI}(\nu_\text{rest,line},z)[W_\kappa(\chi)+W_\text{IA}(\chi)]}{\chi^2}\nonumber\\
    &\times\int \dd{M_h}\,\frac{\dd{n}}{\dd{M_h}}\frac{M_h}{\bar\rho_\mathrm{m}}L_{\text{line}}(M_h)\left|u\left(k=\frac{\ell}{\chi},M_h\right)\right|^2.\label{eq:1hcross}
\end{align}
Unlike with~\autoref{eq:matterbias} we will \emph{not} correct for the fact that $\int \dd{M_h}\,(\dd{n}/\dd{M_h})\,M_h/\bar\rho_\mathrm{m}=1$ does not hold in computational practice, given our imposition of $L(M_h)=0$ for all $M_h<10^{10}h^{-1}\,M_\odot$ for all lines\added{,} which sets the integrand to zero below this limit anyway.

\cite{SchaanWhite21} also consider how the halo density profile introduces scale-dependence into the linear bias for our tracers. In our notation, this turns the constant $\avg{Ib}_\text{line}$ into a scale-dependent
\begin{align}
    \avg{Ib}_\text{line}(k) &= C_\text{LI}(\nu_\text{rest,line},z)\nonumber\\\qquad&\times \int \dd{M_h}\,\frac{\dd{n}}{\dd{M_h}}\,L_\text{line}(M_h,z)b(M_h,z)u(k,M_h).
\end{align}
The matter bias of~\autoref{eq:matterbias} also now becomes scale-dependent:
\begin{equation}
    b_\mathrm{m}(k) = \frac{\int \dd{M_h}\,(\dd{n}/\dd{M_h})\,M_h\,b(M_h)\,u(k,M_h)}{\int \dd{M_h}\,(\dd{n}/\dd{M_h})\,M_h}.
\end{equation}
As these changes impact the LIM--LIM auto and cross spectra as well as the LIM--WL cross spectra, we recalculate these quantities and compare them to the values obtained for the main text.

Since we fix the range of $\ell$ rather than implement scale cuts in $k$, our range of $k$ varies for LIM observables, from 0.3--3 Mpc$^{-1}$ at $z=0.15$ for CO(3--2) at 300 GHz (which we do note is not the dominant source of observed line fluctuations at 300 GHz---cf.~\autoref{fig:observables}), to 0.02--0.2 Mpc$^{-1}$ at $z=8.51$ for \cii{} at 200 GHz. By comparison, the NFW profile only appreciably deviates from 1 in these ranges of $k$ for very high $M_h$---at $z=0.2$, $u(k=3\text{ Mpc}^{-1},M_h)$ barely passes below $0.95$ at $M_h=10^{12}\,M_\odot$ and $0.78$ at $M_h=10^{13}\,M_\odot$. But in addition to the fact that the halo mass function tends to cut off in an exponential-like fashion at high $M_h$, our model works from the SFR model of~\cite{Behroozi19}, which models for a decline in star-formation efficiency at high halo masses. This limits the extent to which $u(k,M_h)$ can affect \emph{any} of the observables outlined.

\begin{figure}
    \centering
    \includegraphics[width=0.96\linewidth]{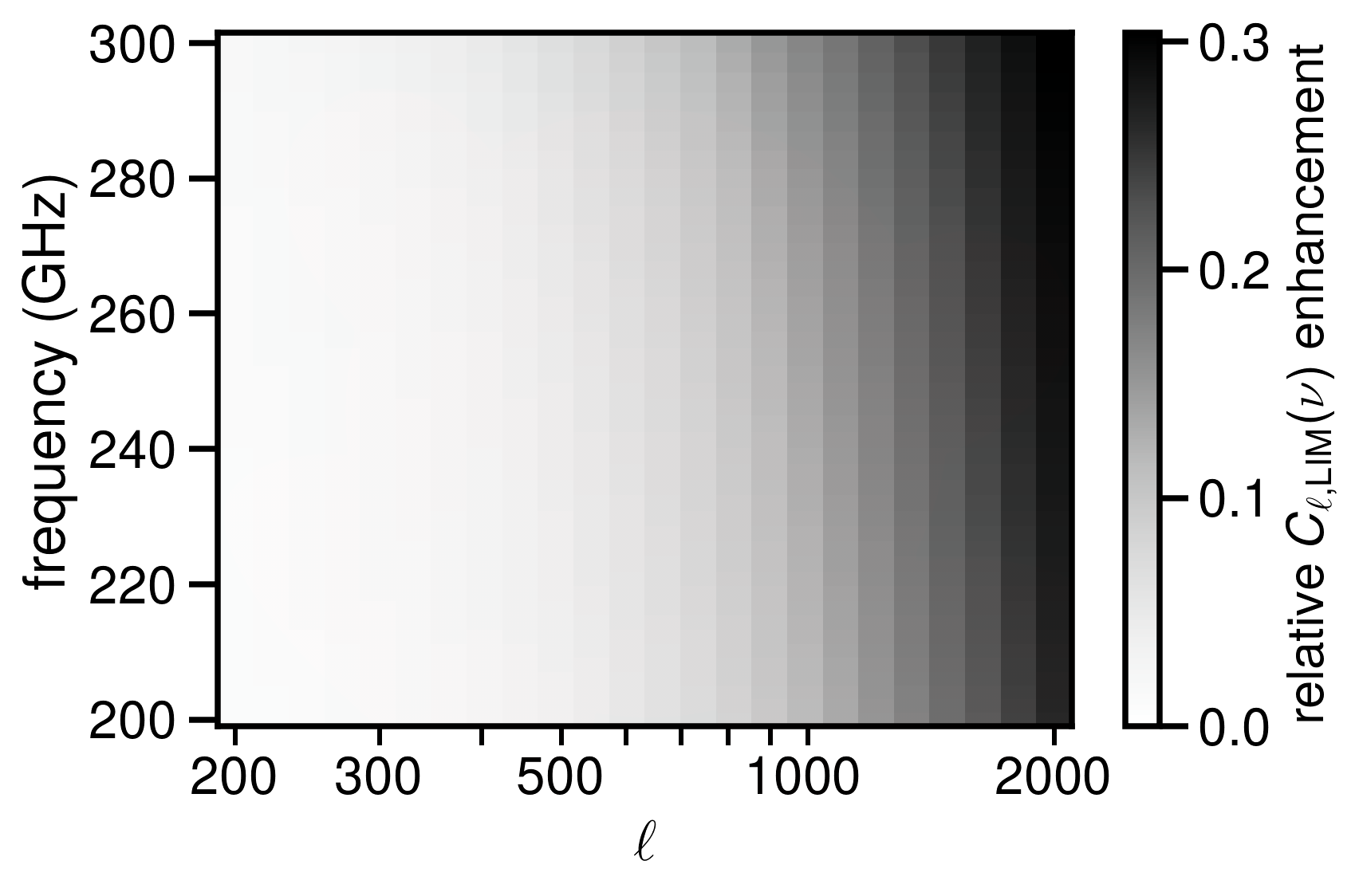}
    \includegraphics[width=0.96\linewidth]{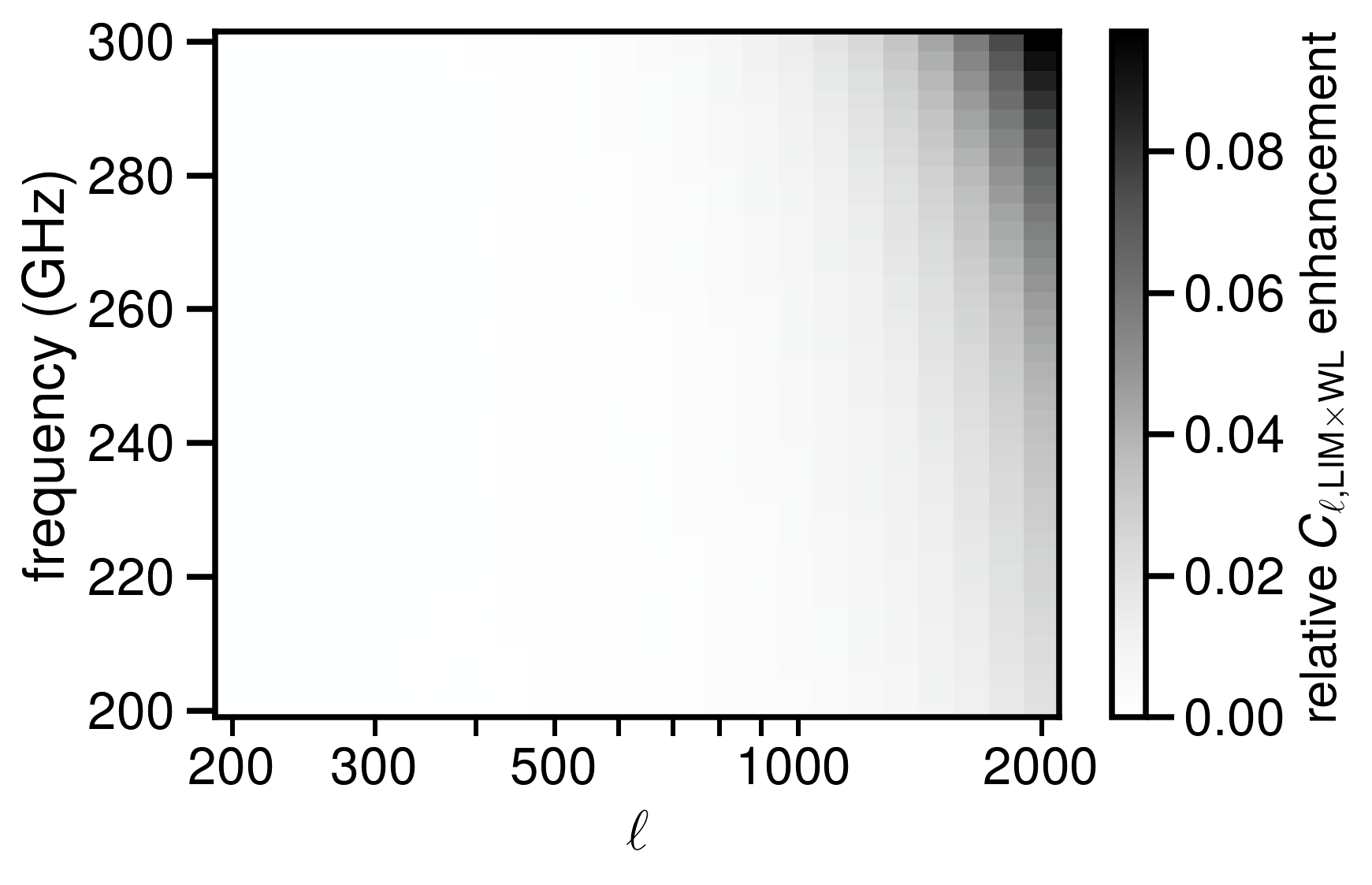}
    \caption{Net relative enhancement from accounting for scale-dependent bias and 1-halo terms of the LIM auto spectra $C_{\ell,\text{LIM}\times\text{LIM}}(\nu,\nu)$ before instrumental noise (upper panel) and the LIM--WL cross spectra $C_{\ell,\text{LIM}\times\text{WL}}(\nu)$ (lower panel), relative to results used for the main text.}
    \label{fig:1hadd}
\end{figure}

To begin with, we find that for our range of $\ell\in(200,2000)$, $\avg{Ib}_\text{line}(k=\ell/\chi)$ is within 1\% of the scale-independent value obtained for all lines at all relevant redshifts except CO(3--2) at $z<0.3$. Continuing by implementing the 1-halo term into the LIM--LIM auto and cross spectra, we find that the resulting difference in calculated values ranges between $+1.0$\% and $+33.6$\% relative to those used for the main text. The LIM--WL cross spectra see a net gain of up to 10\% due to the introduction of the 1-halo term, despite some loss of power due to the introduction of scale-dependent attenuation through the halo profile. \autoref{fig:1hadd} shows the relative enhancements in the cross spectra as well as in the LIM--LIM auto spectra (before the addition of instrumental noise) compared to the main text.

Ultimately, however, we find that the introduction of both the 1-halo term and scale-dependent bias do not significantly alter the forecast signal or its detectability.\addedtwo{ To begin with, a 10\% change in the signal would be significant in a precision cosmology context, but in a mm-wave LIM context where we see greater relative uncertainties in astrophysical modelling (see~\autoref{sec:linemodels}), the credibility of signal forecasts remains far from compromised in ignoring details that result in changes of 10\% at most.} When these details are included in calculation of covariances and detection significances, the resulting total detection significance changes very little in both scenarios: 1.8 with LSST Y1-like $\times$ FYST DSS-like parameters (the same as in the main text), and 48 with LSST Y10-like $\times$ mm-wave Stage 3 parameters (only a few percent down from the value of 50 forecast in the main text). Overall, the $C_{\ell,\text{LIM}\times\text{WL}}$ signal itself has not changed significantly and while the LIM--LIM spectra do see some notable enhancement, it is not enough to substantially alter uncertainties as other scale-independent components like instrumental noise or shape noise likely continue to dominate at high $\ell$. Therefore we expect inferences to also stay largely the same\replaced{ and will not duplicate further work from the main text}{, which we will more explicitly check with a Fisher forecast in~\hyperref[sec:appendix-fisher]{Appendix~\ref{sec:appendix-fisher}}}.

\addedtwo{Note, however, that the above findings do not imply that 1-halo terms or halo density profile effects are unimportant for LIM \emph{in general}. At lower redshift and higher $k$, such small-scale details have far greater impact. In addition, as our understanding of LIM signals and the precision of LIM measurements both improve over time, approximations that suffice now will break down later. Whenever possible, future work should gauge the impact of ignoring or including small-scale details in forecasts, as we have done here.}
    
\section{Fisher forecasts, and comparisons with Monte Carlo-based forecasts in the main text}
\label{sec:appendix-fisher}

We undertake an analysis in the Fisher matrix formalism\footnote{Helpful references include Section 11.4 of~\cite{Dodelson} for a pedagogical overview of Fisher analyses, and~\cite{Coe2009} for further explanation of their use.} to consider potential constraints on the same three parameters considered in the forecasts of~\autoref{sec:mcmc}: $\alpha_\text{IR--CO}$ and $\beta_\text{IR--CO}$, and $A_\text{IA}$.

In the basis of these parameters $\{\lambda_i\}=\{\alpha_\text{IR--CO},\beta_\text{IR--CO},A_\text{IA}\}$, the Fisher matrix is given by
\begin{equation}
    F_{ij} = \sum_{\ell,\ell',\nu,\nu'}\frac{\dd{C}_{\ell,\text{LIM}\times\text{WL}}(\nu)}{\dd\lambda_i}\mathbf{C}^{-1}_{\ell\ell'}(\nu,\nu')\frac{\dd{C}_{\ell',\text{LIM}\times\text{WL}}(\nu')}{\dd\lambda_j},\label{eq:fisher}
\end{equation}
approximating the derivatives with central difference quotients (using a step size of 0.0001 for all parameters). We also modify the Fisher matrix by imposing the same Gaussian prior on $A_\text{IA}$ as in the main text, represented by adding $\sigma_\text{prior}^{-2}[A_\text{IA}]=1$ to the calculation of $F_{A_\text{IA},A_\text{IA}}$ from~\autoref{eq:fisher}. We do not however use the flat prior of $\alpha_\text{IR--CO}\in[0.5,2]$ given in the main text.

\begin{figure}
    \centering
    \includegraphics[width=0.96\linewidth]{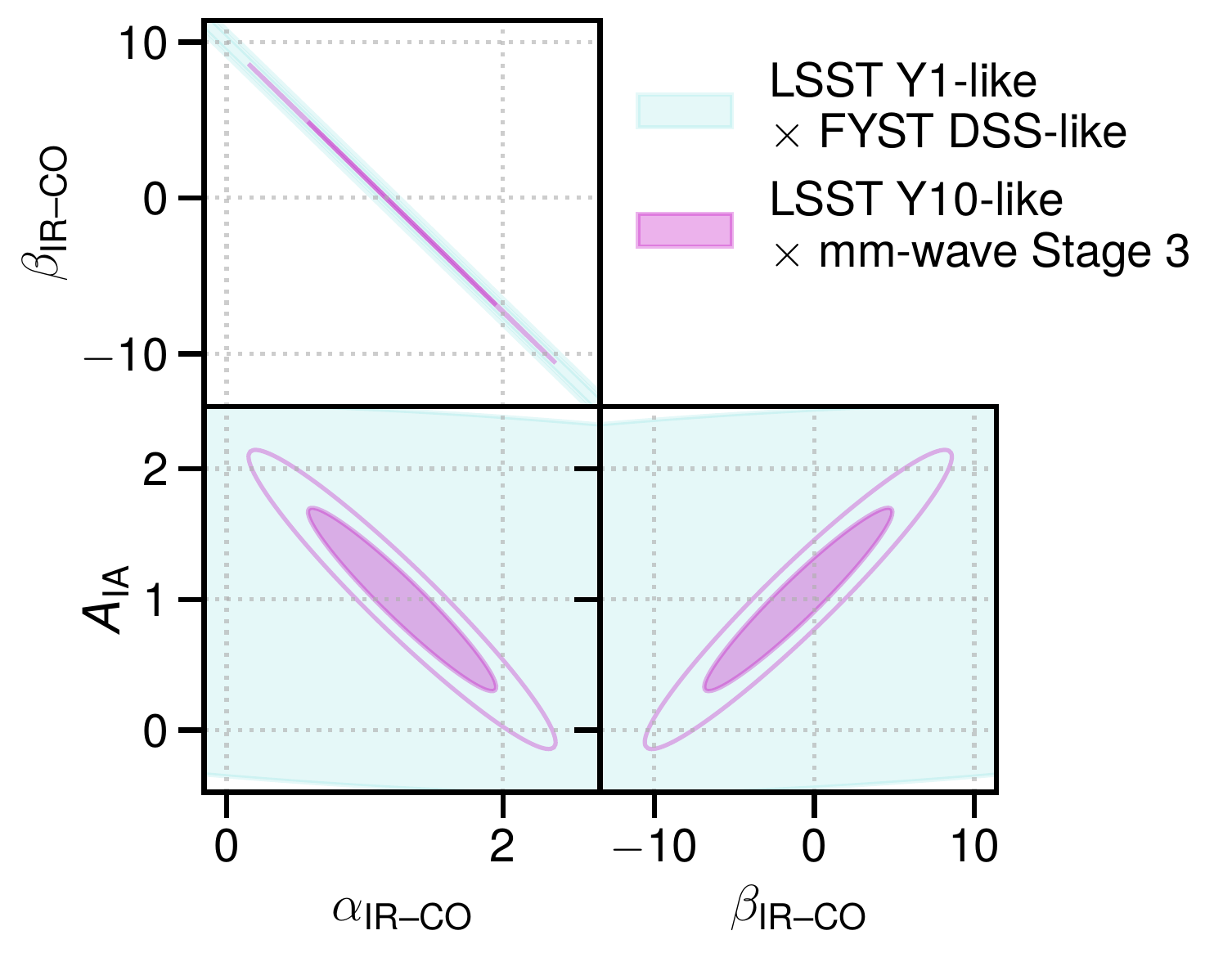}
    \caption{68\% (filled ellipses) and 95\% (unfilled ellipses) confidence regions for the model parameters constrained by our Fisher forecasts given our two experimental scenarios (colours indicated in the legend).}
    \label{fig:fisherel}
\end{figure}

\autoref{fig:fisherel} shows the resulting error ellipses. Much like the detection significance, the constraining power of a LSST Y1-like $\times$ FYST DSS-like cross-correlation is low, with the expected width of the marginalised $A_\text{IA}$ posterior essentially equal to the width of the prior. From this, we know to focus on the constraining power \added{of the} LSST Y10-like $\times$ mm-wave Stage 3 scenario in the main text.

Beyond such broad qualitative conclusions, however, we see that the Fisher forecast fails to capture aspects of the posterior distribution clearly present in the MCMC forecast of~\autoref{sec:mcmc}. Since Fisher forecasts assume Gaussian posteriors, naturally our Fisher forecast cannot predict the heavily skewed distributions estimated in~\autoref{fig:emcee_corner}. Furthermore, while the marginalised posterior of $A_\text{IA}$ itself is not so skewed, we have noted previously that $A_\text{IA}$ is strongly degenerate with the other parameters, whose posteriors are in fact heavily skewed. As a result, the Fisher forecast's posterior estimate for $A_\text{IA}$ of $1.0\pm0.46$ differs somewhat from the MCMC estimate of $A_\text{IA}=0.72^{+0.36}_{-0.33}$. In this case, we do note that the two show the same relative uncertainty, and both agree that the cross-correlation should constrain $A_\text{IA}$ significantly beyond our Gaussian prior. However, the other issues we have discussed motivate using a MCMC forecast for the main text rather than presenting the Fisher forecast as is.

Despite these kinds of prominent issues, Fisher forecasts find ubiquitous use in astrophysical and cosmological contexts, and understandably so given their low computational cost. For this particular work, the Fisher forecast took a few seconds to compute the derivatives and the Fisher matrix, whereas the MCMC presented in~\autoref{sec:mcmc} took 6.26 hours to run all 10000 steps. We may then consider the MCMC to be four orders of magnitude more computationally expensive in terms of `wall time'. But in the grand scheme of all of the work that goes into even a preliminary forecast as presented in this work, a MCMC forecast is decidedly affordable, and will only become more affordable over time as processors become increasingly powerful and efficient. Even now, computers designed for general-purpose use rather than dedicated to high-performance computing can undertake MCMC runs for lower-dimensional parameter spaces in a manageable amount of time.\footnote{All computations in this work used a portable personal computer with a 10-core CPU (Apple M1 Max, with a heterogeneous system of 8 `Firestorm' high-performance cores and 2 `Icestorm' high-efficiency cores).} Therefore, we should perhaps not reflexively choose Fisher forecasts over MCMC runs in situations with skewed or otherwise non-negligibly non-Gaussian parameter probability distributions with strong parameter degeneracies, as in this work. While it is possible to approximately Gaussianise parameter space with parameterised power transforms~\citep{JoachimiTaylor11,Schuhmann16}, this itself requires additional computational effort to determine the transform parameters, which in certain cases may end up being on par with the computational effort required to implement a MCMC forecast.

\added{The Fisher forecast is however useful in certain contexts demanding quick exploratory computation across a number of scenarios, some potentially significantly more computationally expensive to explore, but only needing qualitative conclusions. Already it has allowed us to discard the idea of obtaining any reasonably bounded constraint with the LSST Y1-like $\times$ FYST DSS-like cross-correlation. We now consider a couple of alternate scenarios not considered in the main text to highlight whether or not they are worth investigating further, at least in future work.

Using the Fisher matrix framework, we consider the effect on parameter covariances of adding 1-halo terms and scale-dependent bias as in~\hyperref[sec:appendix-1h]{Appendix~\ref{sec:appendix-1h}}. It is important to establish whether parameter inferences differ significantly as a result of these complications, which introduce a 30-fold increase in the time required to recompute the observable cross spectra at each point in parameter space. (This time could be reduced with further optimisation but it would be premature to expend effort to optimise the calculation before establishing the value of doing so.)

We add 1-halo terms and scale-dependent bias into both signal and covariance calculations and repeat the Fisher matrix analysis, producing the error ellipses shown in the upper panels of~\autoref{fig:fisherel_alts}. We find that the effect on the predicted confidence regions is minuscule, and is dwarfed by the difference between predictions from the MCMC and the Fisher forecast. Given the choice between a model that is marginally more complete but prohibitively expensive for MCMC calculations and a far more MCMC-friendly approximation that is good to within 10\%---with any relative residuals certainly dwarfed by uncertainties around the astrophysics of CO at these redshifts---we opt to use the latter for the main text.

\begin{figure}
    \centering
    \includegraphics[height=0.235\paperheight]{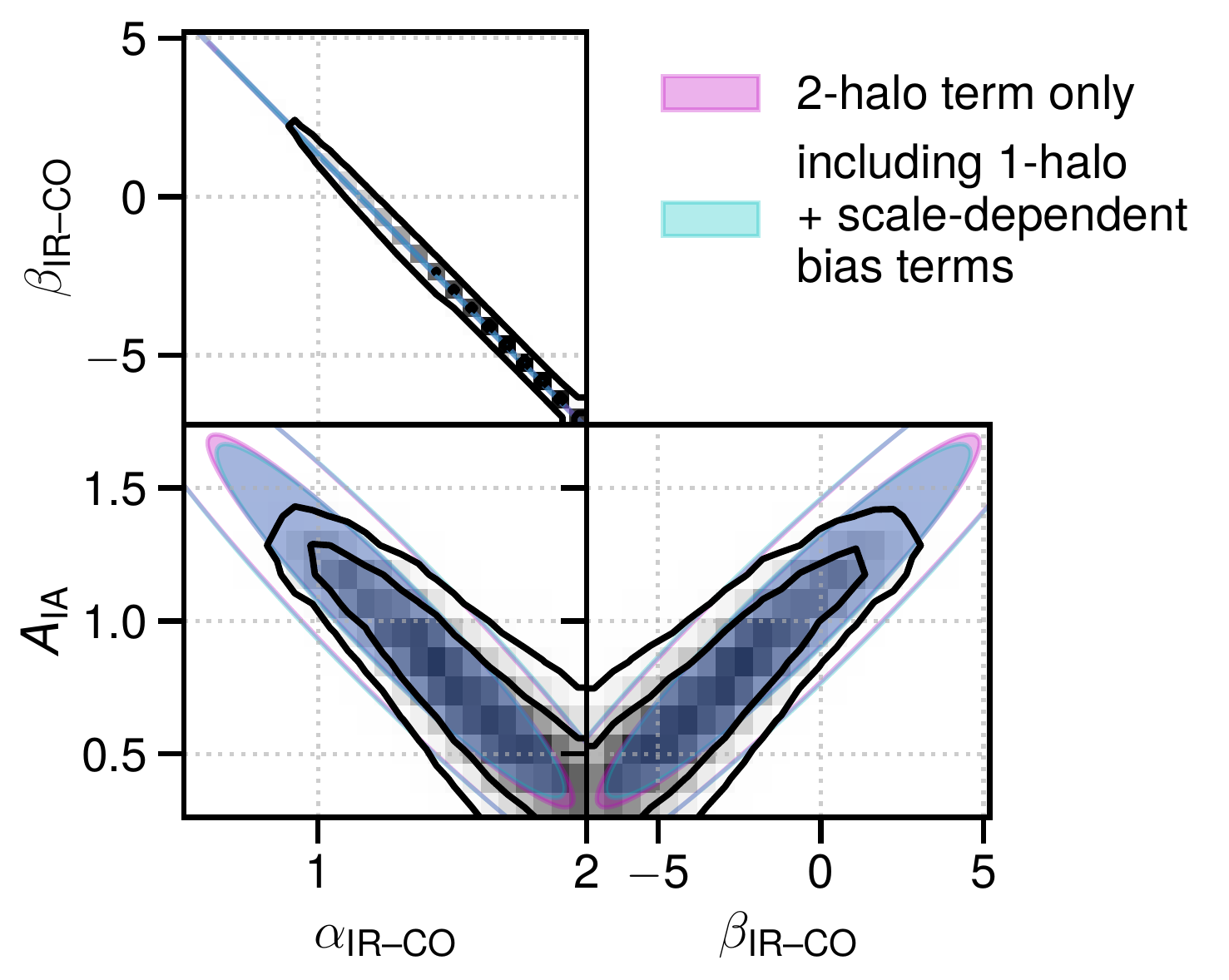}\qquad\includegraphics[height=0.235\paperheight]{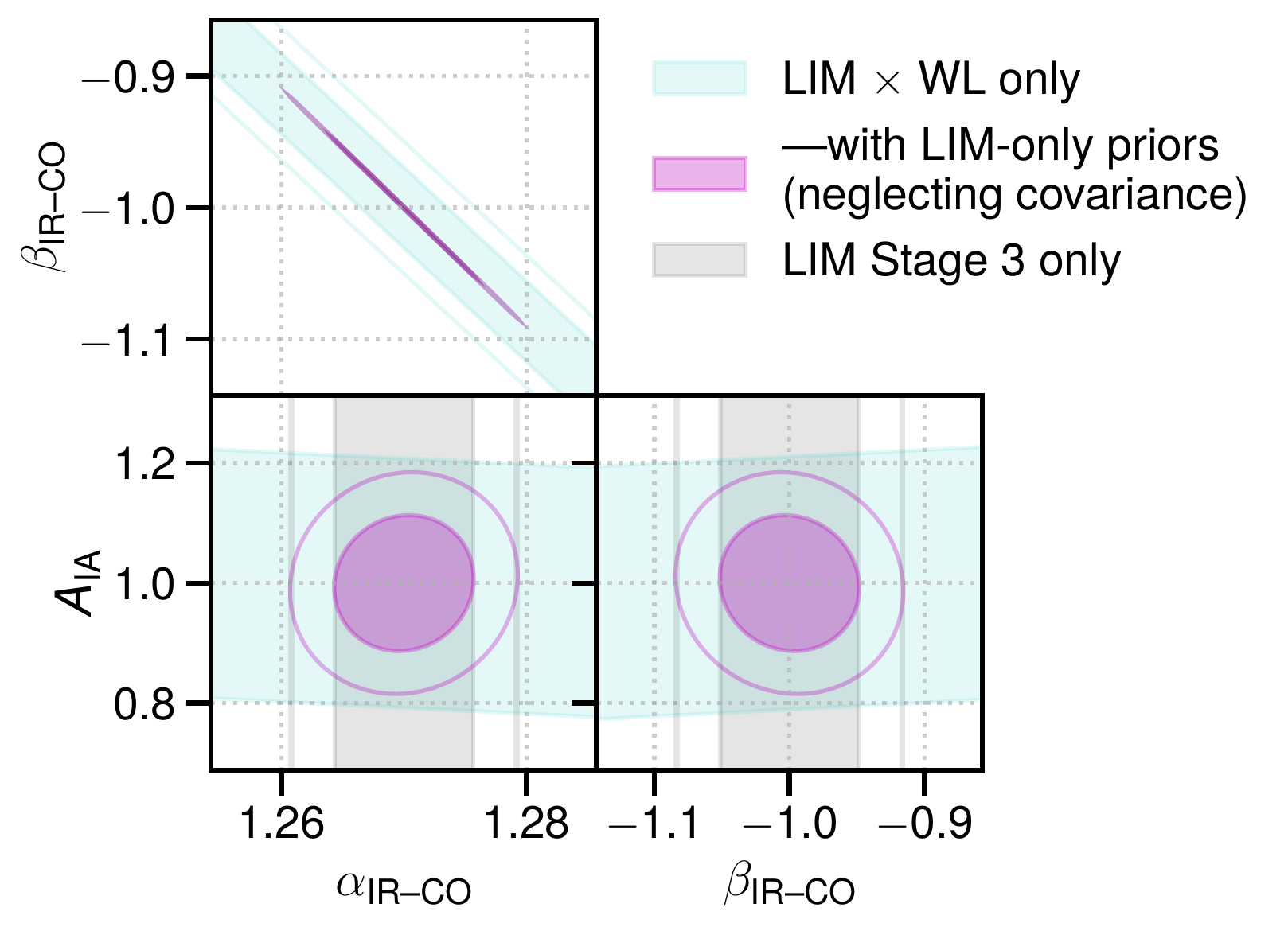}
    \caption{68\% (filled ellipses) and 95\% (unfilled ellipses) confidence regions for the model parameters constrained by additional Fisher forecasts in the LSST Y10-like $\times$ mm-wave Stage 3 scenario, given the inclusion of 1-halo terms and scale-dependent bias (upper panels) or the inclusion of LIM auto $C_\ell$ (lower panels). The upper panels also show the equivalent regions from the MCMC inference in the main text, as shown in~\autoref{fig:emcee_corner}.}
    \label{fig:fisherel_alts}
\end{figure}

We also consider how the LIM auto spectra could improve the constraining power of this measurement, in particular on $A_\text{IA}$. It will be far from surprising for us to find that the combination of auto and cross power spectra breaks parameter degeneracies present in inferences from either one in isolation, given previous forecasts from LIM literature~\citep[e.g.,][, in particular their Appendix D]{COMAPmodels} as well as work in the context of other large-scale structure tracer observations~\citep[e.g.,][]{clearly_a_paper_from_the_referee_probably_at_oxford}.

Here we undertake a highly approximate analysis rather than one that we look to for accurate quantitative answers, especially given that we have already established how parameter confidence regions predicted by Fisher forecasts are unreliable in this context. Specifically, we pretend that the LIM auto spectra and the LIM--WL cross spectra are not correlated at all, which is tenuous but potentially justifiable, for instance by saying that the auto spectra in this case come from an independent measurement of a completely different part of sky.

We calculate the Fisher matrix for the LIM auto $C_\ell$ separately from the LIM--WL cross spectra, taking the covariance between all LIM auto $C_\ell(\nu)$ to be given by
\begin{equation}\mathbf{C}_{\ell\ell',\text{LIM,auto}}(\nu,\nu')=\delta_{\ell\ell'}\cdot\frac{C_{\ell,\text{LIM}\times\text{LIM}}^2(\nu,\nu')}{\ell\,\Delta\ell f_\text{sky}}.\end{equation}
As in the main text, the numerator in this context includes the noise power spectrum as well as the LIM signal.

Using this covariance matrix we find extremely high constraining power on $\alpha_\text{IR--CO}$ and $\beta_\text{IR--CO}$ from the LIM data alone, and if we na\"{i}vely add the LIM auto and LIM--WL cross Fisher matrices together, we find exquisite constraining power on all parameters as shown in the lower panels of~\autoref{fig:fisherel_alts}. In particular, the forecast error on $A_\text{IA}$ decreases from 0.46 to 0.075, a six-fold improvement. This result is important if true, and will need to be revisited more quantitatively in future work with proper accounting of covariances between the auto and cross signals.
}

\bsp	
\label{lastpage}
\end{document}